\documentclass[12pt]{article}

\usepackage{amsmath}
\usepackage{amssymb}

\usepackage{graphicx}

\usepackage{cite}

\usepackage{color} 

\usepackage{setspace} 
\usepackage[labelfont=bf,labelsep=period,justification=raggedright]{caption}

\makeatletter
\renewcommand{\@biblabel}[1]{\quad#1.}
\makeatother

\date{}

\begin{document}

\begin{flushleft}
{\Large
\textbf{Robustness of circadian clocks to daylight fluctuations: hints
  from the picoeucaryote Ostreococcus tauri}
}
\\
Quentin Thommen$^{1,2,3,4}$,
Benjamin Pfeuty$^{1,2,3,4}$, 
Pierre-Emmanuel Morant$^{1,2,3,4}$, 
Florence Corellou$^{5,6}$,
Fran\c{c}ois-Yves Bouget$^{5,6}$,
Marc Lefranc$^{1,2,3,4,\ast}$
\\
\bf{1} Universit\'e Lille 1, Laboratoire de
    Physique des Lasers, 
    Atomes, et Mol\'ecules, UFR de Physique,
    F-59655 Villeneuve d'Ascq, France
\\
\bf{2} CNRS, UMR 8523, CERLA, FR2416,
  F-59655 Villeneuve d'Ascq Cedex, France
\\
\bf{3} Universit\'e Lille
  1, Institut de Recherche Interdisciplinaire, F-59655 
    Villeneuve d'Ascq, France
\\
\bf{4} CNRS, USR 3078,
  F-59655     Villeneuve d'Ascq, France
\\
\bf{5} Universit\'e Pierre and Marie Curie Paris
    06, Laboratoire
    d'Oc\'eanographie Microbienne, Observatoire
    Oc\'eanologique, F-66651 Banyuls/Mer, France
\\
\bf{6} Centre National de la Recherche Scientifique, Laboratoire 
           d'Oc\'eanographie Microbienne, Observatoire Oc\'eanologique, 
           F-66651 Banyuls/Mer, France
\\
$\ast$ E-mail: marc.lefranc@univ-lille1.fr
\end{flushleft}


\section*{Abstract}

    The development of systemic approaches in biology has put emphasis
    on identifying genetic modules whose behavior can be modeled
    accurately so as to gain insight into their structure and
    function. However most gene circuits in a cell are under control
    of external signals and thus quantitative agreement
    between experimental data and a mathematical model is difficult.
    Circadian biology has been one notable exception: quantitative
    models of the internal clock that orchestrates biological
    processes over the 24-hour diurnal cycle have been constructed for
    a few organisms, from cyanobacteria to plants and mammals. In most
    cases, a complex architecture with interlocked feedback loops
    has been evidenced. Here we present first modeling results for the
    circadian clock of the green unicellular alga 
    \emph{Ostreococcus tauri}. Two plant-like clock genes have been
    shown to play a central role in \emph{Ostreococcus} clock. We find that
    their expression time profiles can be accurately reproduced
    by a minimal model of a two-gene transcriptional feedback loop.
    Remarkably, best adjustment of data recorded under light/dark
    alternation is obtained when assuming that the
    oscillator is not coupled to the diurnal cycle. This suggests
    that coupling to 
    light is confined to specific time intervals and has no dynamical
    effect when the 
    oscillator is entrained by the diurnal cycle. This intringuing
    property may reflect a strategy to minimize the impact of
    fluctuations in daylight intensity on the core circadian
    oscillator, a type of perturbation that has been rarely
    considered when assessing the robustness of circadian clocks.

\section*{Author Summary}

Circadian clocks keep time of day in many living organisms, allowing
them to anticipate environmental changes induced by day/night
alternation. They consist of networks of genes and proteins
interacting so as to generate biochemical oscillations with a period
close to 24 hours. Circadian clocks synchronize to the day/night cycle
through the year principally by sensing ambient light. Depending on
the weather, the perceived light intensity can display large
fluctuations within the day and from day to day, potentially inducing
unwanted resetting of the clock. Furthermore, marine organisms such as
microalgae are subjected to dramatic changes in light intensities in
the water column due to streams and wind. We showed, using
mathematical modelling that the green unicellular marine alga
Ostreococcus tauri has evolved a simple but effective strategy to
shield the circadian clock from daylight fluctuations by localizing
coupling to the light during specific time intervals. In our model, as
in experiments, coupling is invisible when the clock is in phase with
the day/night cycle but resets the clock when it is out of phase. Such
a clock architecture is immune to strong daylight fluctuation.

\section*{Introduction}
Real-time monitoring of gene activity now allow us to
unravel the complex dynamical behavior of regulatory networks
underlying cell functions~\cite{shav-tal04:_imagin}. However,
understanding the collective behavior of even a few molecular actors
defies intuition, as it depends not only on the topology of the
interaction network but also on strengths and response times of
its links~\cite{alon06:_introd_system_biolog}. A mathematical
description of a regulatory network is thus necessary to qualitatively and
quantitatively understand its dynamical behavior, but obtaining it is 
challenging. State variables and parameters are subject to large
fluctuations~\cite{elowitz02:_stoch}, which create artificial
complexity and mask the actual network structure. Genetic modules
are usually not isolated but coupled to a larger network, and a given
gene can be involved in different modules and
pathways~\cite{LouiseAshall04102009}. It is thus important to identify
gene circuits whose dynamical behavior can be modeled quantitatively,
to serve as model circuits. 

One strategy for obtaining such circuits has been to construct
synthetic networks, which are isolated by
design~\cite{Elowitz00:_synth_net,gardner00:_switch,Stricker08:_synth_osc}. 
As recent experiments have shown, an excellent quantitative agreement
can be obtained by incorporating when needed detailed descriptions of
various biochemical processes (e.g., multimerization, transport, DNA
looping, etc.)~\cite{Stricker08:_synth_osc}.

Another strategy is to study natural gene circuits whose function
makes them relatively autonomous and stable. The circadian clocks that
drive biological processes around the day/night cycle in many living
organisms are natural candidates, as these genetic oscillators keep
track of the most regular environmental constraint: the alternation of
daylight and darkness caused by Earth
rotation~\cite{panda02:_circad,Goldbeter96book,Ouyang98:_circa_osc,Dodd05:_circa_osc}.
Informed by experiments, circadian clock models have progressively
become more complex, evolving from single loops featuring a
self-repressed gene~\cite{Goldbeter95a,Leloup99a} to networks of
interlocked feedback
loops\cite{forger:_mammalian,locke06:_exper,zeilinger06:_arabid_prr7_prr9,francois05:_neurospora}.

Here we report surprisingly good agreement between the mathematical
model of a single transcriptional feedback loop and expression
profiles of two central clock genes of \emph{Ostreococcus tauri}. This
microscopic green alga is the smallest free-living eukaryote known to
date and belongs to the Prasinophyceae, one of the most ancient groups
of the green lineage. \emph{Ostreococcus} displays a very simple
cellular organization, with only one mitochondrion and one
chloroplast~\cite{Courties94:_ostreo_struc2,chretiennot95:_ostreo_struc}.
Its small genome (12.6 Mbp) sequence revealed a high compaction (85\%
of coding DNA) and a very low gene redundancy~\cite{derelle06:_genome}
(e.g., most cyclins and CDK are present as a single copy
gene~\cite{moulager07:_cell_divis_ostreoc}).The cell division cycle of
\emph{Ostreococcus} is under control of a circadian oscillator, with
cell division occurring at the end of the day in light/dark
cycles~\cite{moulager07:_cell_divis_ostreoc}. These daily rhythms in
cell division meet the criteria characterizing a circadian clock, as
they can be entrained to different photoperiods, persist under
constant conditions and respond to light pulses by phase shifts that
depend on internal time~\cite{moulager07:_cell_divis_ostreoc}.

Very recently, some light has been shed on the molecular workings of
\emph{Ostreococcus} clock by Corellou \emph{et al.}
\cite{corellou09:_toc1_cca1}. Since the clock of closely related
\emph{Arabidopsis} has been extensively studied, they searched
\emph{Ostreococcus} genome for orthologs of higher plant clock genes
and found only two, similar to \emph{Arabidopsis} central clock genes
\emph{Toc1} and \emph{Cca1}~\cite{corellou09:_toc1_cca1}. These two
genes display rhythmic expression both under light/dark alternation
and in constant light conditions. A functional analysis by
overexpression/antisense strategy showed that \emph{Toc1} and
\emph{Cca1} are important clock genes in \emph{Ostreococcus}.
Overexpression of \emph{Toc1} led to increased levels of CCA1 while
overexpression of \emph{Cca1} resulted in lower levels of TOC1.
Furthermore CCA1 was shown to bind to a conserved evening element
sequence (EE) that is required for the circadian regulated activity of
\emph{Toc1} promoter. Whether \emph{Toc1} and \emph{Cca1} work in a
negative feedback loop could not be inferred from this study since
\emph{Ostreococcus} clock appeared to rely on more than a simple
\emph{Toc1}/\emph{Cca1} negative feedback loop.

Interestingly, \emph{Arabidopsis} genes \emph{Toc1} and \emph{Cca1}
were the core actors of the first plant clock model, based on a
transcriptional loop where TOC1 activates \emph{Cca1} and the
similar gene \emph{Lhy}, whose proteins dimerize to
repress \emph{Toc1}~\cite{DavidAlabadi08032001,locke05:_modelling}.
However, this model 
did not reproduce well expression peaks of \emph{Toc1} and
\emph{Cca1} in \emph{Arabidopsis}~\cite{locke05:_modelling}
and was extended to adjust experimental
data~\cite{locke05:_extension}. Current \emph{Arabidopsis} clock
models feature several interlocked feedback
loops~\cite{locke06:_exper,zeilinger06:_arabid_prr7_prr9}. This led us
to investigate whether the transcriptional feedback loop model where
\emph{Toc1} activates \emph{Cca1} and is repressed by \emph{Cca1}
would be relevant for \emph{Ostreococcus}.

We not only found that this two-gene loop model reproduces perfectly
transcript profiles of \emph{Ostreococcus} \emph{Toc1} and \emph{Cca1}
but that excellent adjustment of data recorded under light/dark
alternation is obtained when no model parameter depends on light
intensity. This counterintuitive finding suggests that the oscillator
is not permanently coupled to light across the 24-hour cycle but only
during specific time intervals, which is supported by numerical
simulations. In this article, we propose that the invisibility of
coupling in entrainment conditions reflects a strategy to shield the
oscillator from natural fluctuations in daylight intensity.

\section*{Results}
\subsection*{Experimental data and model adjustment}
To characterize the temporal pattern of \emph{Toc1} and \emph{Cca1}
expression in \emph{Ostreococcus}, we used microarray data acquired in
triplicate under 12:12 light/dark cycle, as described
in~\cite{moulager07:_cell_divis_ostreoc} (Fig.~\ref{fig:data}). One
\emph{Toc1} and two \emph{Cca1} mRNA time courses had no aberrant
point. Here, we use as target profiles the complete \emph{Toc1}
profile and the complete \emph{Cca1} profile whose samples are
obtained from the same microarray data as the \emph{Toc1} profile. We
checked that the results 
described in this work are robust to the biological variations
observed. Corellou \emph{et al.} have also carried out an extensive
work of genetic transformation in \emph{Ostreococcus}, leading to
transcriptional and translational fusion lines allowing one to monitor
transcriptional activity and protein dynamics in living
cells~\cite{corellou09:_toc1_cca1}. However, luciferase kinetics in
this organism is still not well known and we postpone the analysis of
luminescence time series to a future work. Model adjustment has thus
been carried out using microarray expression data, which reflect
accurately the endogeneous levels of mRNA. Although seeking
quantitative agreement with luminescence time series was premature at
this stage, predicted protein concentration profiles were compared
with data from translational fusion lines as an additional test.

A minimal mathematical model of the two-gene feedback loop comprises
four ordinary differential equations (Eq.~(\ref{eq:model}), Methods)
with 16 parameters.
Since detailed models extending the basic 4-ODE model~(\ref{eq:model})
could only have led to better adjustment, we purposely neglected here
effects such as compartmentalisation or delays due to transcription or
translation so as to minimize the risk of overfitting and reliably
assess the validity of the two-gene loop hypothesis.

Experimental data are recorded under 12:12 Light/Dark (LD) alternation
so that the coupling which synchronizes the clock to the diurnal cycle
must be hypothesized. Circadian models usually assume that some
parameters depend on light intensity (e.g., a degradation rate
increases in the dark), and thus take different values at day and
night. Parameter space dimension then increases by the number of
modulated parameters. Various couplings to light were considered, with
1 to 16 parameters depending on light intensity. We also tested
adjustment to model~\eqref{eq:model} with all parameters constant,
which allowed us to quantify the relevance of coupling mechanisms
by measuring the difference between best-fitting profiles in the
coupled and uncoupled cases.

The free-running period (FRP) of the oscillator in constant day
conditions was fixed at 24 hours, which was the mean value observed in
experiments~\cite{corellou09:_toc1_cca1}, but we checked that our main
results remain valid for other values of the FRP. In fact, we found
that when FRP was freely adjustable, it usually converged to values
close to or slightly below 24 hours. Fixing the FRP at exactly 24
hours is interesting in that coupling mechanisms are selected by
adjustment only if they improve goodness of fit and not merely to
achieve frequency locking.

\subsection*{A free-running model adjusts experimental data}
\label{sec:freerunning}
The first result is that an excellent agreement between numerical and
experimental profiles is obtained, with a root mean square (RMS) error
of a few percent (Figs.~\ref{fig:adjustment}(A)-(B)). There is no
point in extending model~\eqref{eq:model} to improve adjustment of
microarray data, which are compatible with the hypothesis of a
\emph{Toc1}-\emph{Cca1} feedback loop. Moreover, the corresponding
protein profiles (not adjusted) correlate well with luminescence
signals from CCA1:Luc and TOC1:Luc translational fusion lines
(Figs.~\ref{fig:adjustment}(C)-(F)).

But the more surprising is that a non-coupled model, where all
parameters are kept constant, adjusts experimental data
(Fig.~\ref{fig:adjustment}(B), RMS error 3.6\%) essentially as well as
a fully coupled model where all parameters are allowed to vary between
day and night (Fig.~\ref{fig:adjustment}(A), RMS error 3.3\%). The
corresponding parameter values are given in
Table~\ref{tab:parameters}. When only one or a few parameters were
modulated, goodness of fit significantly degraded compared to the
uncoupled and fully coupled cases. This indicates that besides being
biologically unrealistic, the model with all parameters modulated fits
data merely because of its large parameter space dimension, and cannot
be considered seriously. Moreover we simulated the transition from LD
alternation to constant light (LL) or constant darkness (DD)
conditions for this model and found that it 
still adjusted experimental data well in LL while displaying strongly
damped oscillations in DD (Fig. S1). This confirms that adjustment
relies on time profiles being close to free-running oscillator
profiles and that adjustment by a fully coupled model is in fact accidental.

On the other hand the uncoupled model is equally unrealistic because
it cannot be entrained to the day/night cycle, whereas it is observed
experimentally that upon a phase shift of the light/dark cycle, CCA1
and TOC1 expression peaks quickly recover their original timings in
the cycle. To verify that adjustment by a free-running oscillator
model does not depend on the target profile used, we generated a large
number of synthetic profiles whose samples where randomly chosen
inside the interval of variation observed in biological triplicates,
and adjusted a free-running oscillator model to them. In each case, we
found that although RMS error slightly degraded compared our target
profile (where mCCA1 and mTOC1 samples for a given time always come
from the same microarray), it remained on average near 10 \%, with visually
excellent adjustment (Fig. S2). Last, it should be noted that assuming
a FRP of 24 hours allows frequency locking to occur without coupling,
but cannot induce best adjustment in this limiting case by itself.

Thus the paradoxical result that data points fall
almost perfectly on the temporal profiles of a free-running oscillator
is counterintuitive but must nevertheless be viewed as a signature of
the clock architecture. As we will see, this in fact does not imply
that the oscillator is uncoupled but only that within the class of
models considered so far, where parameters of the TOC1--CCA1 loop take
day and night values, the uncoupled model is the one approaching
experimental data best. Nothing precludes that there are more general
coupling schemes that adjust data equally well.

Before unveiling such models, we discuss now whether the
simple negative feedback loop described by model~\eqref{eq:model} is a
plausible autonomous gene oscillator. With two transcriptional
regulations, it is a simpler circuit than the Repressilator, where
three genes repress themselves circularly~\cite{Elowitz00:_synth_net}.
It is known that in this topology, oscillations become more stable as
the number of genes along the loop increases. The two-gene feedback
loop described by~(\ref{eq:model}) could therefore seem to be a less
robust oscillator than the Repressilator, and thus a poor model for
the core oscillator of a circadian clock.

\begin{figure}[!ht]
  \centering
  \includegraphics[width=15cm]{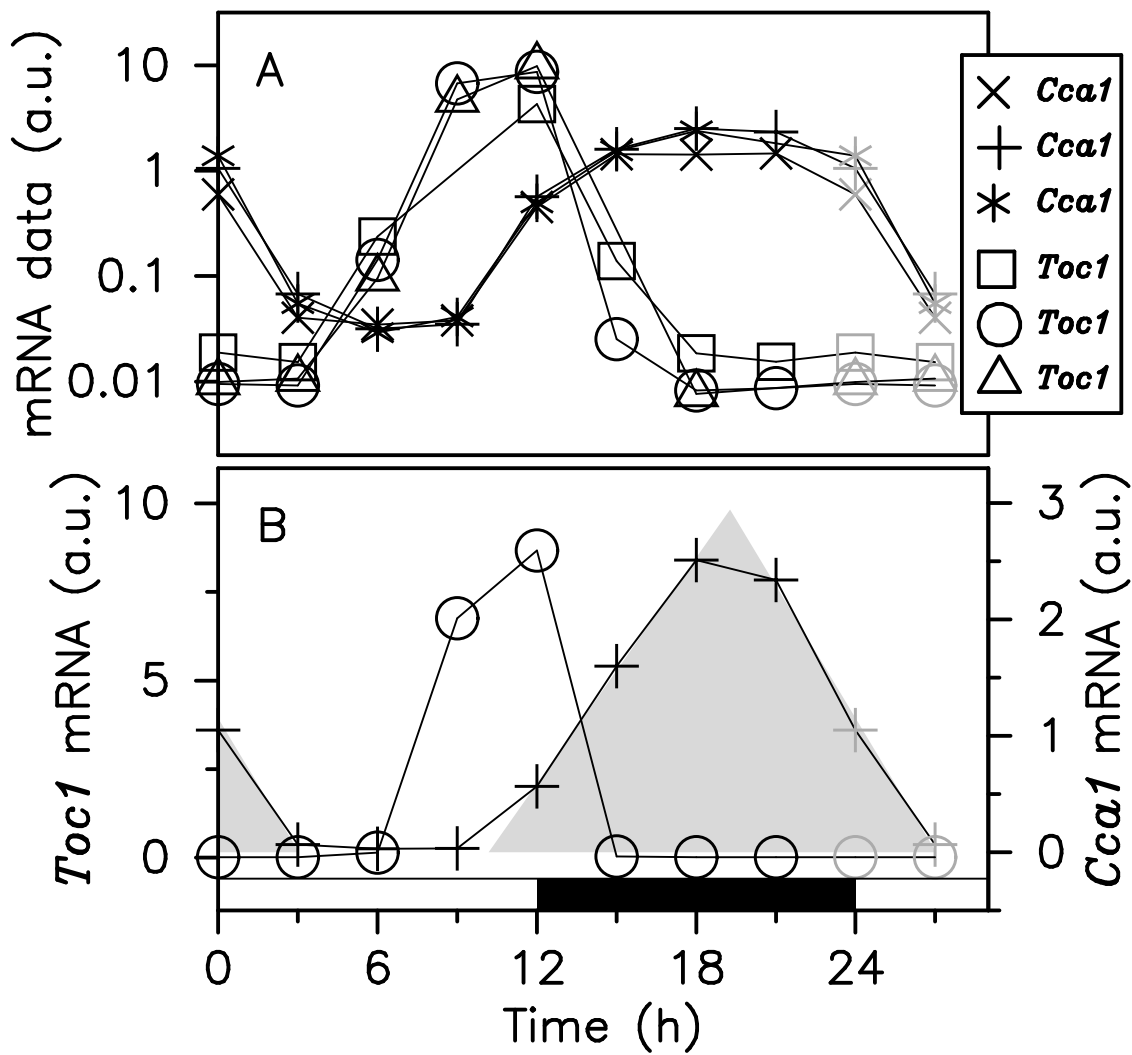}

  \caption{\textbf{Microarray data recorded under 12:12 LD
      alternation}. Time zero corresponds to dawn. (A) Experimental
    data points for the \emph{Cca1} and \emph{Toc1} mRNA time profiles
    are drawn in logarithmic scale. Data points at zeitgeber time (ZT)
    0 and ZT3 have been replicated in gray at ZT24 and ZT27. The
    target \emph{Toc1} and \emph{Cca1} profiles selected for
    subsequent analysis are shown with circles and pluses,
    respectively. These two profiles are also shown in linear scale in
    (B), where the shaded area illustrates the sawtooth shape of the
    \emph{Cca1} mRNA profile, which will be used later as evidence of
   a strongly saturated enzymatic degradation. This area has been
   obtained by fitting a 
    straight line through \emph{Cca1} data points at ZT12, ZT15 and
    ZT18 on one hand and at ZT21, ZT0 and ZT3 on the other hand. }
  \label{fig:data}
\end{figure}

\begin{figure}[!ht]
  \centering
  \includegraphics[width=15cm]{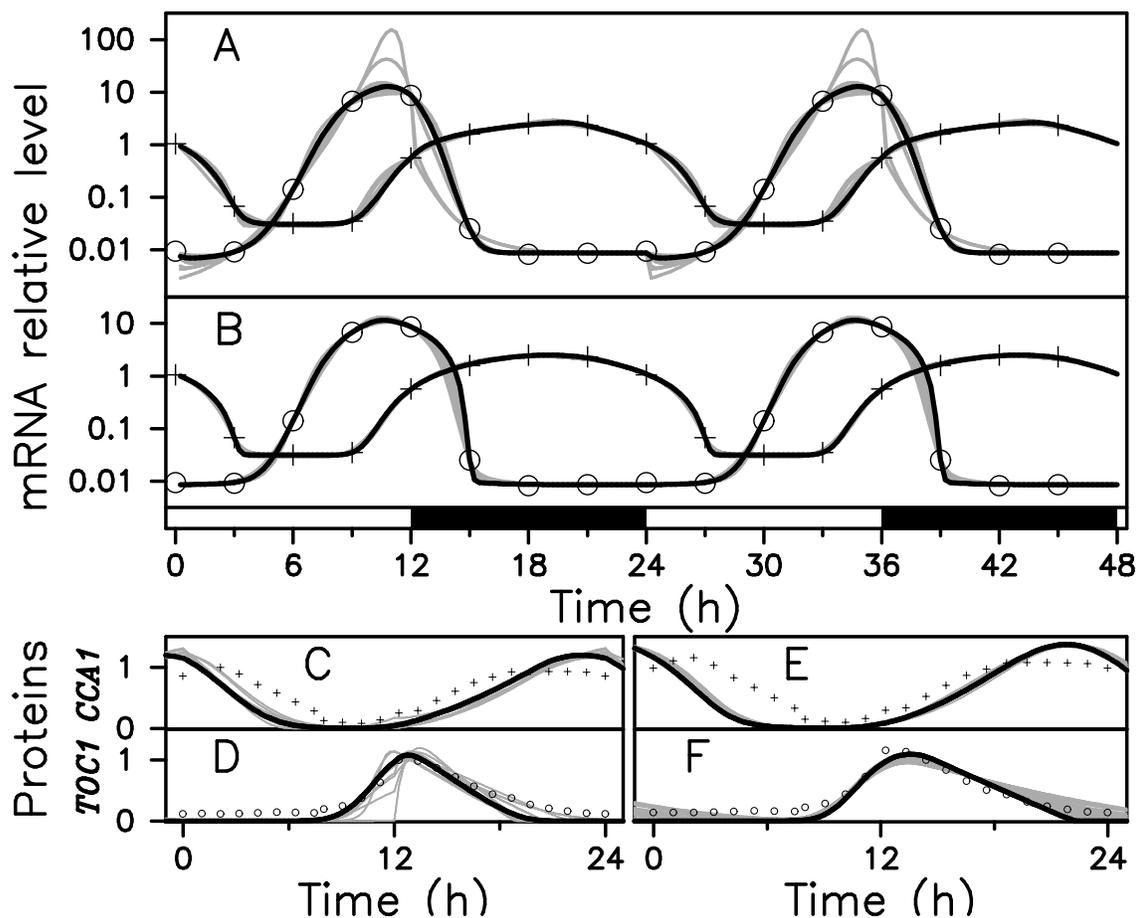}
  \caption{\textbf{Adjustment of experimental data}. The data of
    Fig.~\ref{fig:data}(A) are adjusted by model~(\ref{eq:model}) with
    a FRP of 24 hours. In (A) and (B), crosses (resp. circles)
    indicate the \emph{Cca1} (resp. \emph{Toc1}) microarray data used
    as target. Solid lines are best-fitting mRNA time profiles (log
    scale) obtained with models where (A) all parameters are coupled
    to light; (B) no parameter is coupled to light; a few solutions
    near optimum are shown in gray with the best one in black. (C)
    (resp. (E)) solid lines are CCA1 predicted time profile (linear
    scale) corresponding to (A) (resp. (B)) with the same color code;
    crosses correspond to luminescence signals from translational
    fusion lines. (D) and (F) are the same curves as (C) and (E) for
    the TOC1 protein.}
  \label{fig:adjustment}
\end{figure}

To address this issue, we checked robustness of adjustment with
respect to parameter variations. We found that the experimental
profiles can be reproduced in a wide region of parameter space around
the optimum, which is quite remarkable given the simplicity of the
model (Fig. S3). Moreover, a distinctive feature of the best fitting
parameter sets is a strongly saturated degradation, in particular for
\emph{Cca1} mRNA, with an extremely low value of $K_{M_C}$ equal to
$0.6 \%$ of the maximal \emph{CCa1} mRNA concentration (see
table~\ref{tab:parameters}). In this situation, the number of
molecules degraded per unit time is essentially constant and does not
depend on the concentration except at very small values. This is
consistent with the characteristic sawtooth shape of our target
profile drawn in linear scale (Fig.~\ref{fig:data}(B)).

The role of post-translational interactions in gene oscillators and
circadian clocks has been recently emphasized (see,
e.g.,~\cite{MFL,Tomita05:_cyanobac}), and in particular saturated
degradation has since long been known to favor
oscillations~\cite{Goldbeter96book,tiana07:_oscil,kurosawa02:_satur}.
Recently, it has been been shown to act as a delay~\cite{degradator,
  Mather:_tsimringPRL} and to be essential for inducing robust
oscillations in simple synthetic
oscillators~\cite{Wong07:_prot_deg,Stricker08:_synth_osc,cookson09}
(compare Fig.~\ref{fig:data}(B) with Fig. 5 of \cite{cookson09}).
Thus, strongly saturated degradation is very 
likely also a key 
dynamical ingredient of the natural gene oscillator studied here.

\subsection*{Adjustment by a model with gated coupling}

Circadian models are usually coupled to diurnal cycle by changing some
parameter values between day and
night~\cite{Goldbeter95a,Leloup99a,forger:_mammalian,locke06:_exper,zeilinger06:_arabid_prr7_prr9,francois05:_neurospora}.
This assumes that all molecular actors involved in light input
pathways have been incorporated and that their properties (e.g.,
degradation rates) react directly to light. Such couplings act over
the entire cycle except when light-sensitive actors are present only
transiently. For example, models of \emph{Arabidopsis} clock feature
an intermediary protein PIF3 that is necessary for induction of CCA1
by light but is shortly degraded after dawn so that CCA1 transcription
is only transiently
activated\cite{locke05:_modelling,locke05:_extension,locke06:_exper}.
Gating of light input has been observed in several circadian clocks
and may be important for maintaining proper timing under different
photoperiods~\cite{FlorianGeier02012005}.

In our case, light/dark alternation has no detectable signature in the
dynamics of \emph{Toc1} and \emph{Cca1} mRNA when the clock is
phase-locked to the diurnal cycle. This suggests that the actors of
the two-gene loop do not sense light directly, and are driven via unknown
mediators, which modify their 
properties inside specific temporal intervals.  Since the input pathway can
have complex structure and dynamics, possibly featuring separate
feedback loops, the windows of active coupling may be located anywhere
inside the diurnal cycle and reflect light level at other times of the
cycle. Coupling activation should depend both on time of day and on
the intrinsic dynamics of the light input pathway, notwithstanding a
possible feedback from the circadian core
oscillator~\cite{bognar99:_phytochrome,RekaToth12012001,salazar09:_predic}.

For simplicity, we restrict ourselves to models in which some
parameters of the TOC1--CCA1 feedback loop are modified between two
times of the day, measured relatively to dawn (ZT0). The start and end
times of coupling windows are then model parameters instead of
being fixed at light/dark transitions. This assumes that the input
pathway tracks diurnal cycle instantaneously, without loss of
generality for understanding behavior in entrainment conditions. In
this scheme, resetting of the two-gene oscillator can be studied by
simply shifting the oscillator phase relatively to the coupling
windows. The results so obtained will be sufficient to show that there
exist coupling schemes which leave no signature on mRNA profiles, and
to study their properties.

What makes our approach original is not the gated coupling to diurnal
cycle, which can be found in other models, but the fact that we do not
try to model the actors of the input pathway, which can be complex.
This is because we focus here on the TOC1--CCA1 feedback loop, which
mostly behaves as an autonomous oscillator. Thus we only need to know
the action of the unknown mediators on TOC1 or CCA1, the details of
their dynamics being irrelevant.

\begin{figure}[!ht]
  \centering
  \includegraphics[width=15cm]{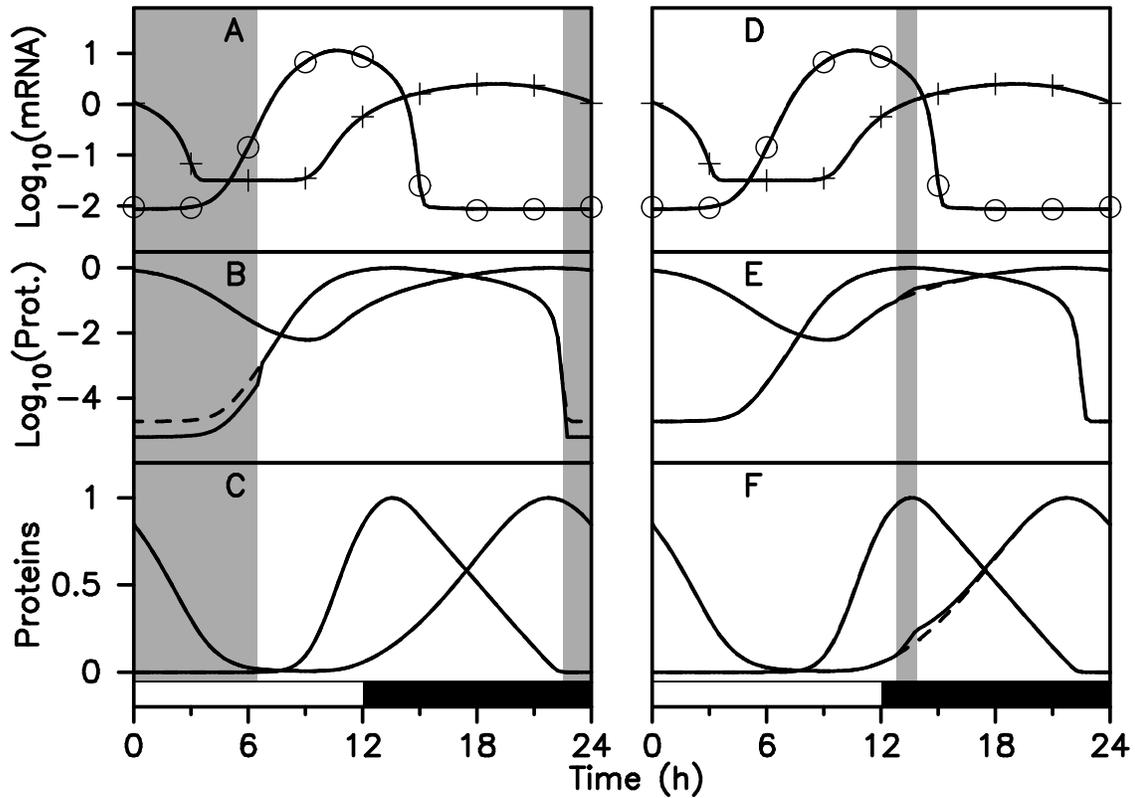}
  \caption{\textbf{Adjustment by models with gated coupling}.
    Numerical solutions of model~(\ref{eq:model}) without coupling
    (dashed lines, same parameter values as in
    Fig~\ref{fig:adjustment}(B)) and with coupling (solid lines). Gray
    areas indicate coupling activation. In the left (resp. right)
    column, TOC1 (resp. CCA1) degradation rate is multiplied by 3
    (resp. divided by 2) from ZT22.5 to ZT6.5 (resp. from ZT12.8 to
    ZT13.95). (A), (D) mRNA time profiles; protein time profiles are
    shown in (B), (E) logarithmic scale and (C), (F) linear scale.}
  \label{fig:occas_coupling}
\end{figure}

We systematically scanned the coupling window start and end times,
adjusting model for each pair. This revealed that many coupling
schemes are compatible with experimental data. For example, TOC1
degradation rate $\delta_{P_T}$ can be modified almost arbitrarily in
a large temporal window between ZT22.5 and ZT6.5 without degrading
adjustment. This is shown in Figs.~\ref{fig:occas_coupling}(A)-(C),
where $\delta_{P_T}=3\delta^0_{P_T}$ inside this window (here and
below, $\delta^0_X$ denotes the uncoupled degradation rate of variable
$X$). Although the coupling is active for 8 hours, this coupling
scheme generates mRNA and protein profiles which are indistinguishable
from those of a free-running oscillator. Indeed, modifying TOC1
stability in a window where protein level is low, as is the case for
any subinterval of the ZT22.5--ZT6.5 window, does not perturb the
oscillator.

We also found a family of time windows of different lengths centered
around ZT13.33, inside which the CCA1 degradation rate $\delta_{P_C}$
can be decreased without significantly modifying goodness of fit. In
Figs.~\ref{fig:occas_coupling}(D)-(F), we show the effect of having
$\delta_{P_C}=\delta^0_{P_C}/2$ between ZT12.8 and ZT13.95. In this
coupling scheme, mRNA profiles are not affected but coupling
activation has a noticeable effect on CCA1 level, which rises faster
than in the uncoupled case. After the window, however, CCA1 level
relaxes in a few hours to the uncoupled profile, losing memory of the
perturbation. Near this time of the day, the CCA1 protein level
appears to be slaved by the other variables: the perturbation induced
by modified degradation does not propagate to the other variables, and
when coupling is switched off, the protein level relaxes to its value
in the uncoupled solution. Thus, the effect of coupling is not only
small but transient. An important consequence, which we will exploit
later, is that the two coupling windows shown in
Fig.~\ref{fig:occas_coupling} can be combined without modifying
adjustment, provided the perturbation induced by one window has vanished
when the other window begins.

In these examples, adjustment is sensitive to the timing of these
coupling windows: when the start time is modified slightly, the end
time must be changed simultaneously so as to recover good adjustment.
On the other hand, we found that adjustment error depends little on
the coupling strength (measured by the ratio between degradation rates
outside and inside the window), especially for short coupling windows.

\begin{figure}[htbp]
  \centering
    \includegraphics[width=15cm]{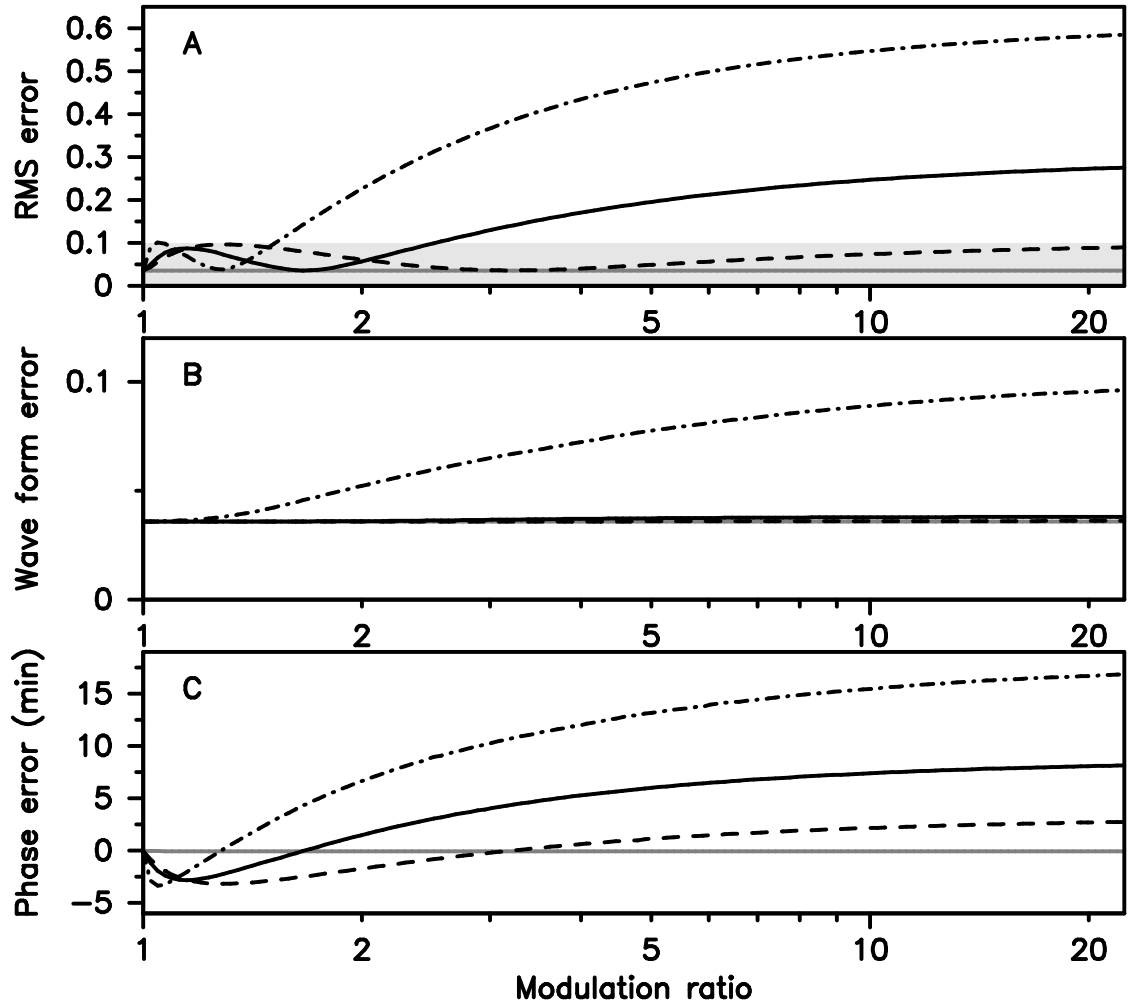}
    \caption{\textbf{Adjustment error as a function of the coupling
        amplitude for three coupling windows}. (A) In gray, RMS error
      when $\delta_{P_T}$ is multiplied by $r$ from ZT22.5 to ZT6.5;
      in black RMS error when $\delta_{P_c}$ is divided by $r$ from
      ZT12.8 to ZT13.95 (solid), from ZT13 to ZT13.65 (dashed), and
      from ZT12 to ZT15.47 (dash-dotted). The shaded area correspond
      to adjustment RMS errors below  10 \%. (B) Waveform error, given by
      the minimal adjustment error obtained when a variable phase
      shift is applied to the entrained oscillations; (C) Phase error,
      defined as the phase shift for which the minimal adjustment
      error is obtained.}
  \label{fig:rmserror}
\end{figure}

Fig.~\ref{fig:rmserror}(A) shows how adjustment error varies as a
function of coupling strength for the two coupling windows used in
Fig.~\ref{fig:occas_coupling} as well as for two other windows inside
which the CCA1 protein degradation is reduced, one shorter and the
other longer than the window in Fig.~\ref{fig:occas_coupling}(B). The
window of accelated TOC1 degradation is totally insentitive to
modifications of the TOC1 degradation rate, which is due to protein
levels being very low in this window. Windows of CCA1 stabilization
are all the more insensitive to variations in CCA1 degradation rate as
they are shorter. To quantify the sensitivity of a given window we
define $r_\text{max}$ as the largest value of the ratio
$r=\delta^0_{P_C}/\delta_{P_C}$ such that adjustment RMS error remains
below 10 \% for any value of $r$ between $1$ and $r_\text{max}$. The
associated variations in mRNA profiles are visually undetectable and
below experimental uncertainties. For the windows ZT12--ZT15.47,
ZT12.8--ZT13.95 and ZT13--ZT13.65, of respective durations 3.47, 1.15
and 0.65 hours, we find that the $r_\text{max}$ index takes the value
1.5, 2.5 and 260 respectively.

To gain better insight into the effect of a coupling window, we must
take into account the fact that the induced variation in the entrained
oscillations can be decomposed as a displacement along the limit cycle
(resulting in a phase shift) and a displacement transversely to the
limit cycle (resulting in a deformation of the limit cycle). To this
end, we apply a variable phase shift to the entrained time profile and
optimize this phase shift so as to minimize the adjustment error. We
define the waveform error as the minimal value of the latter, and the
phase error the value of the phase shift for which it is obtained. A
small waveform error indicates that we are following the same limit
cycle as in the free-running case, possibly with a different phase
than is observed experimentally.
Waveform and phase errors for the three windows of CCA1 protein
stabilization considered in Fig.~\ref{fig:rmserror}(A) are shown in
Figs.~\ref{fig:rmserror}(B) and \ref{fig:rmserror}(C), respectively.
It can be seen that only the largest window is associated with a
deformation of the limit cycle for large values of $r$, and that it
remains modest (RMS error of about 10 \% for $r=20$). For the two
shorter windows, degraded adjustment essentially results from a phase
shift of the entrained solution as the modulation index is increased.
It can also 
be seen that the phase error is in fact very small, approximately 7.5
and 2.5 minutes at $r=10$ for the two shorter windows. Thus 
it appears that for short enough windows, the effect of the light
coupling mechanism can be entirely captured by studing the phase response
induced by the mechanism and that a necessary property of a coupling
window is that it induces a zero phase shift of the free-running limit
cycle (or a phase shift corresponding to the mismatch between the
natural and forcing periods in the general case that we will consider
later).

\subsection*{Systematic characterization of gated coupling mechanisms}

Besides the two specific examples shown in
Fig.~\ref{fig:occas_coupling}, other coupling schemes are compatible
with experimental data. In this section, we undergo a systematic
approach in order to determine those coupling schemes that do
synchronize the free-running model to the day/night cycle, while
leaving no signature on mRNA profiles when the phase-locking regime is
achieved. To this aim, a preliminary step is to identify those
coupling schemes that synchronize in the limit of weak forcing using
the tools of infinitesimal phase response curve, which can be
defined in the framework of perturbation theory in the vicinity of
periodic
orbits~\cite{Guckenheimer83book,kramer84:_sensit,rand04:_desig}.
Computation of the parametric impulse phase response curve
~\cite{Taylor08:_sensit} ($Z_{piPRC}$)
characterizing a light-coupling mechanism corresponding to parameter
variation $\mathbf{dp}$ allows one to
determine time intervals specified by duration $\tau$ and median position
$t_m$ such that when the mechanism is applied in this time interval, it
generates a zero phase shift and phase-locking is stable to small
perturbations (see Supporting materials). Such intervals satisfy:
\begin{equation}
\left\{ \begin{array}{rll}
\int_{t_m-\tau/2}^{t_m+\tau/2} \, Z_{piPRC}(u,\mathbf{dp})\, du=0 \\
\int_{t_m-\tau/2}^{t_m+\tau/2} \, Z_{piPRC}'(u,\mathbf{dp})\, du<0
\end{array} \right.
\label{Eq:z}
\end{equation}

\begin{figure}[!ht]
  \begin{center}
    \includegraphics[width=11cm]{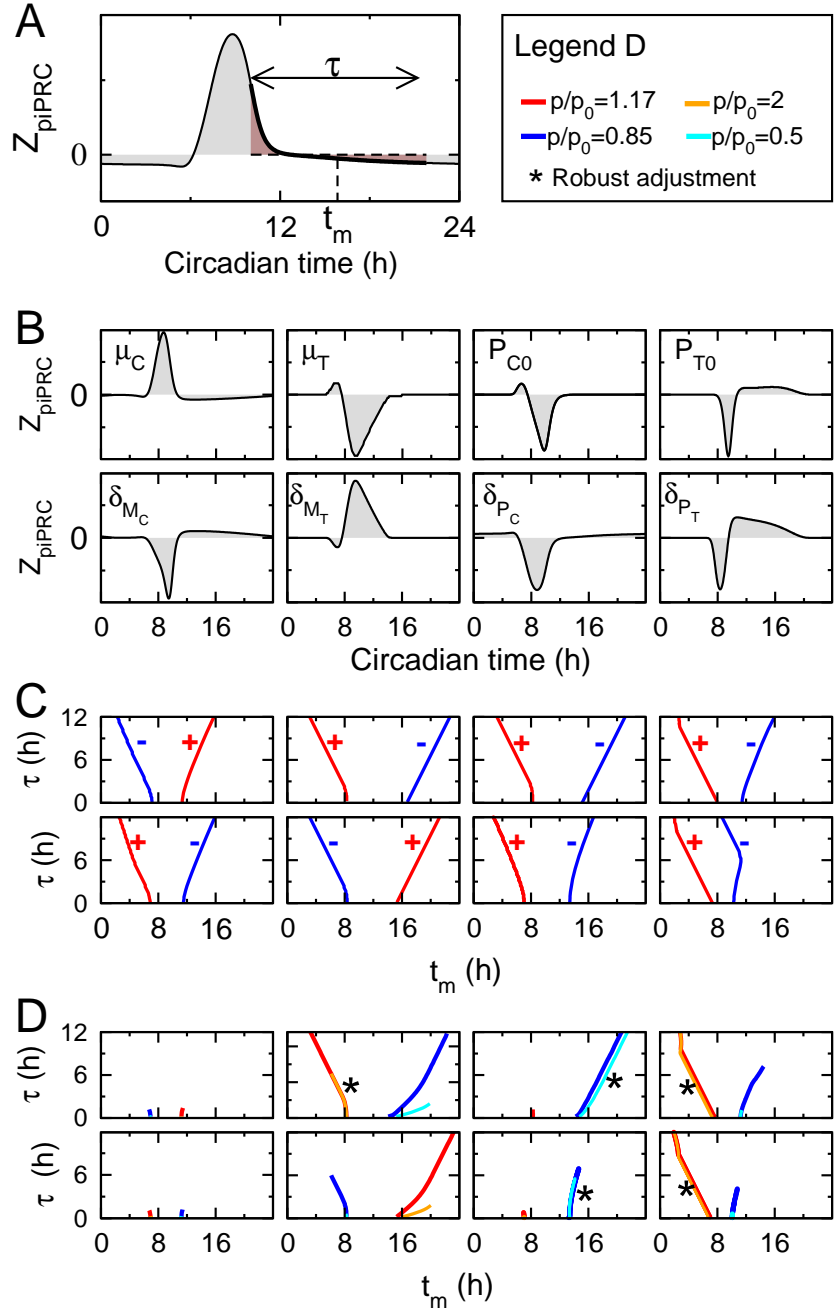}
  \end{center}
  \caption{\textbf{Characterization of coupling schemes.} (A)
    Schematic representation of how window center and duration $t_m$
    and $\tau$, which characterize a coupling with rectangular gating
    profile, are estimated from the piPRC using Eq.~(\ref{Eq:z}). (B)
    piPRC characterizing the phase change induced by an infinitesimal
    perturbation of some parameters of the model (transcription an
    degradation kinetics). (C) Characterization of window center $t_m$
    and duration $\tau$ satisfying Eq.~(\ref{Eq:z}) for the coupling
    mechanisms shown in (B), as illustrated in (A). Parameters chosen
    in (B) are modulated either positively (red) or negatively (blue).
    (D) Characterization of window center and duration of gated
    couplings which adjust experimental data with a RMS error below $10\%$ 
    for two different coupling strengths (see box on the right-hand side of
    the top: 
    $p/p_0$ is the ratio between the parameter values within and outside the  
    coupling window).}  
  \label{fig:syst}
\end{figure}

Figure \ref{fig:syst} depicts the properties of various gated
couplings in the case where the light-coupling mechanism is assumed to
modulate specifically a single transcription-related or
degradation-related kinetic parameter. For sufficiently weak positive
or negative modulation of those eight parameters, a coupling window of
specific width ($\tau$) and position ($t_m$) can always be found to
satisfy the Eq. \ref{Eq:z} (Figs. \ref{fig:syst}(A)--(C)), thus being
compatible with experimental data. However, the adjustment of these
weak coupling schemes to data is expected to deteriorate progressively
when coupling strength is increased, because (i) the locking phase may
change, (ii) the modulation may deviate significantly the trajectory
from that of the free-running oscillator or (iii) the entrained
solution may loose its stability. Numerical simulations performed at
different coupling strengths indicate that only a subset of coupling
schemes determined in the limit of weak coupling keep a good
adjustement irrespective of the coupling strength. Fig.
\ref{fig:syst}(D) shows window timings such that adjustment error
remains below 10 \% when the kinetic parameter is multiplied or
divided by 1.17 or 2. Such a goodness of fit can only be obtained if
limit cycle deformation remains small.

As with the examples considered in the previous section, some coupling
mechanisms have robust adjustment properties in that a good adjustment
is obtained at the two different coupling strengths for the same
timings, which coincide with the timings computed in the weak coupling
limit. In these cases, adjustment is robust to variations in the
coupling strength, which suggests that for these coupling mechanisms,
the weak coupling approximation 
remains valid up to large coupling strengths. For instance, light
coupling mechanisms that temporarily increase TOC protein degradation
($\delta_{P_T}$) or CCA1 activation threshold ($P_{T0}$)
in windows located during the day the night appear to be robust couplings. 
Similarly, decreasing CCA1 protein degradation ($\delta_{P_C}$) or TOC
repression threshold ($P_{C0}$) in windows occuring during night are robust
light-coupling mechanisms.
Some other mechanisms do not display the same robustness because
either the window timings corresponding to good adjustment depend
sensitively on coupling strength (e.g., for positive modulation of
mTOC1 degradation rate) or because no good adjustment can be found
except for very short windows (e.g., modulation of mCCA1 degradation
rate). Other robust coupling mechanisms can be identified in Fig.
S4, in which the coupling mechanisms not considered in
Fig.~\ref{fig:syst} are characterized.

\begin{figure}[!ht]
  \begin{center}
    \includegraphics[width=14cm]{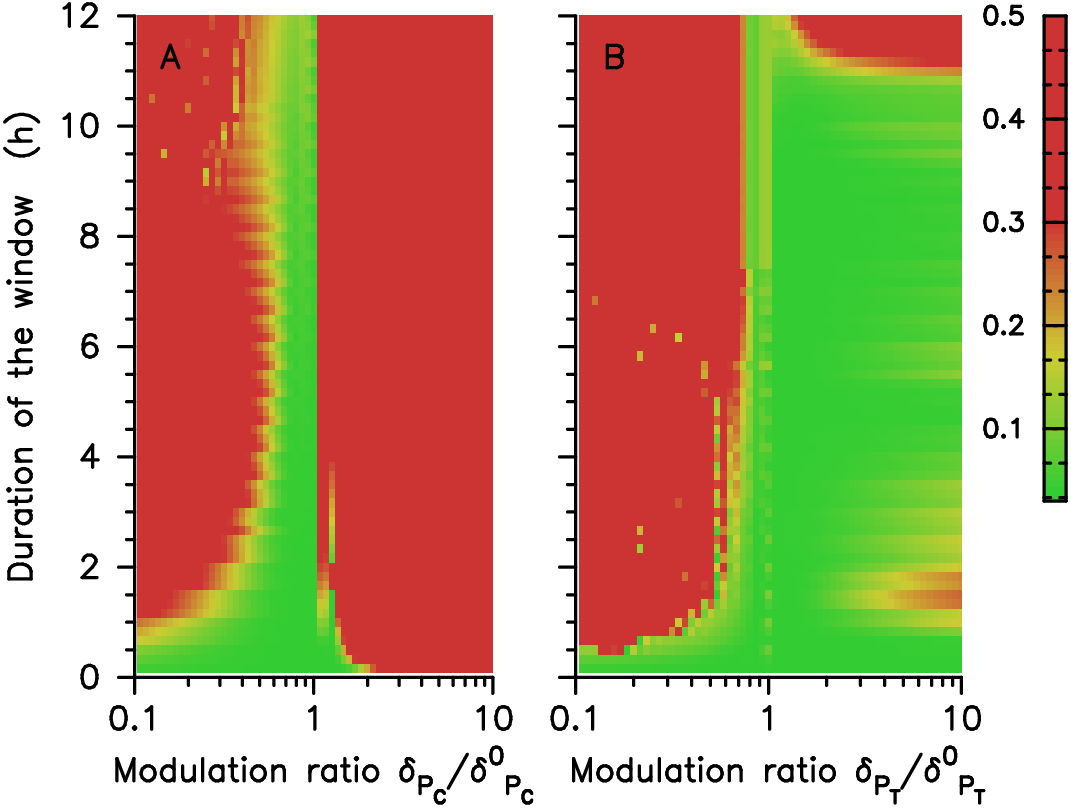}
  \end{center}
  \caption{\textbf{Robustness of adjustment with respect to coupling
      strength and window duration.} Color-coded adjustment RMS error
    as a function of window duration and modulation ratio (ratio of
    degradation rates inside and outside the coupling window). (A)
    Modulation of CCA1 protein degradation rate; (B) Modulation of
    TOC1 protein degradation rate.}
  \label{fig:greenred}
\end{figure}

Figure~\ref{fig:greenred} provides a complementary illustration of the
robustness of adjustment for models with gated modulation of CCA1 or
TOC1 protein degradation rate. In these plots, the window center is
kept fixed at the time determined from Eq.~(\ref{Eq:z}) and shown in
Fig.~\ref{fig:syst}(C) while coupling strength and window duration are
freely varied. It can be seen that this timing is compatible with
adjustment in a wide range of coupling strengths and window durations.

Our analysis shows that several coupling mechanisms are compatible
with the experimental data and that discriminating them requires
more experimental data. In particular, monitoring gene expression in
transient conditions will probably be crucial since the coupling
mechanism leaves apparently no signature in the experimental data in
entrainement conditions. For simplicity, we restrict ourselves in the
following to models in which half-lives of TOC1 or CCA1 proteins are
modified during a specific time interval that is determined in Fig
\ref{fig:syst}(D).

\subsection*{Resetting}
\begin{figure}[!ht]
  \centering
  \includegraphics[width=15cm]{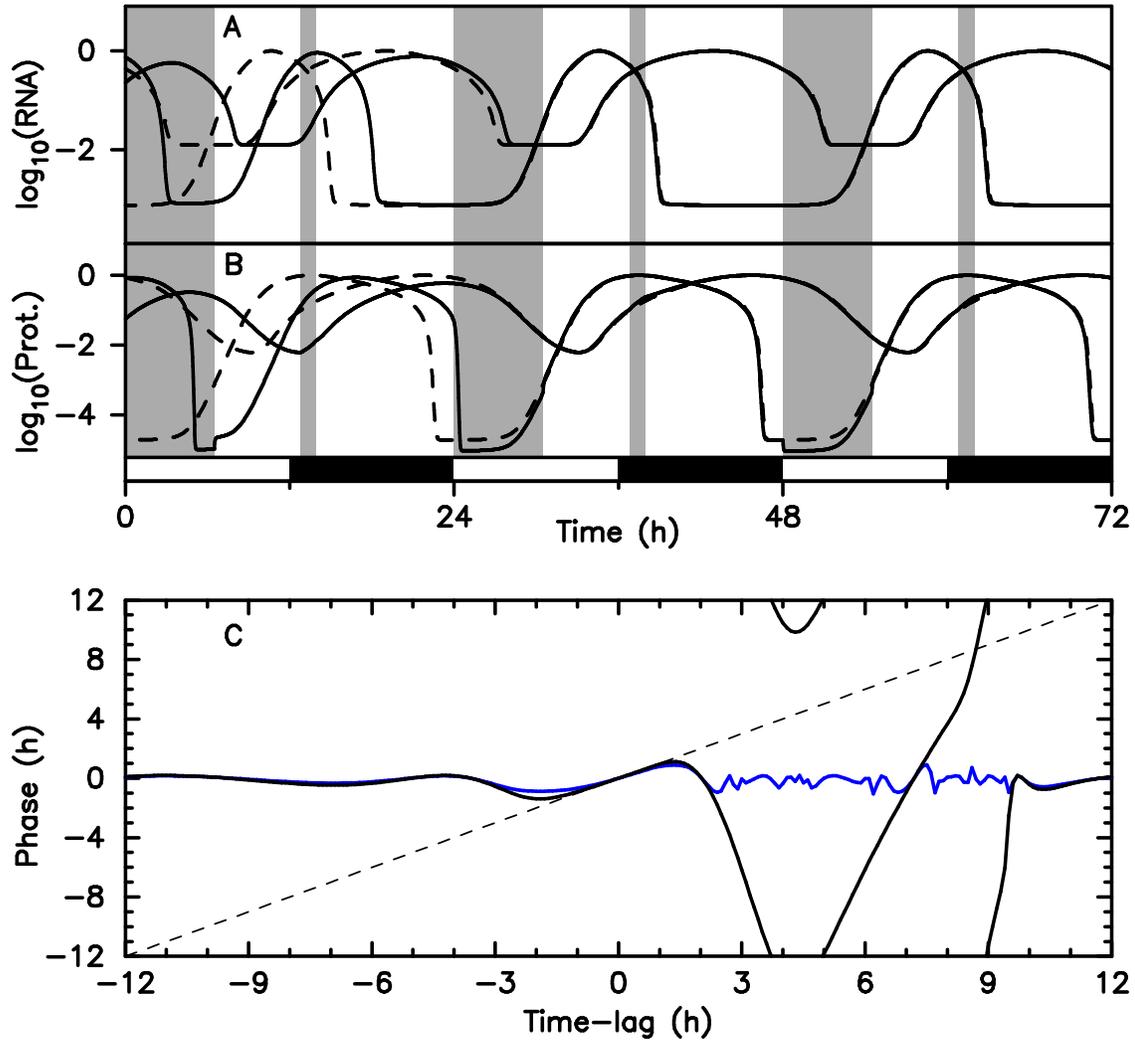}
  \caption{\textbf{Resetting properties of a model with gated
      coupling}. TOC1 (resp. CCA1) degradation rate is multiplied by
    2.1 (resp., by 0.6) from ZT0 to ZT6.5 (resp., from ZT12.8 to
    ZT13.95). After phase-shifting the day/night cycle by 12 hours,
    (A) mRNA and (B) protein time profiles (logarithmic scale) of
    numerical solutions (solid lines) converge rapidly to the nominal
    profile (dashed lines). (C) Residual phase shift one day (black)
    and five days (blue) after a phase shift ranging from -12 to 12
    hours has been applied.}
  \label{fig:shiftrecover}
\end{figure}

One may wonder about the purpose of coupling schemes with almost no
effect on the oscillator. The key point is that our data have been
recorded when the clock was entrained by the diurnal cycle and
phase-locked to it. A natural question then is: how do such couplings
behave when clock is out of phase and resetting is needed? We found
that while the two mechanisms shown in Fig.~\ref{fig:occas_coupling}
have poor resetting properties when applied separately (Fig. S5), a
combination of both can be very effective. In
Fig.~\ref{fig:shiftrecover}(A)-(B), we show how the two-gene
oscillator recovers from a sudden phase-shift of 12 hours using a
two-window coupling scheme. As described above, we assume for
simplicity that the two coupling windows remain fixed with respect to
the day/night cycle. The 12-hour phase shift is induced by
initializing at dawn the oscillator state with the value it takes at
dusk in the entrained regime. Figs.~\ref{fig:shiftrecover}(A)-(B)
show that most of the lag is absorbed in the first 24 hours and the
effect of the initial perturbation is hardly detectable after 48
hours.

To design this coupling, we utilized the fact that modifying coupling
strengths inside windows hardly affects adjustment. We could therefore
choose their values so as to minimize the maximal residual phase shift
after three days for all possible initial lags
(Fig.~\ref{fig:shiftrecover}(C)). Interestingly, we found that the
best resetting behavior is obtained when the start time of the window
of modified TOC degradation coincides with dawn. Phase locking in this
example is globally stable. However, resetting becomes slow when the
residual phase shift is under an hour and the residual phase shift is
variable (RMS phase error after 5 days is 25 minutes and maximum phase
error is 1 hour), and (Fig.~\ref{fig:shiftrecover}(C)). This
inefficiency results in fact from the limitations of a model where the
two parameters are modulated by a rectangular profile with fixed
timing. Indeed, we will see later that impressive adjustment and
resetting behavior can be simultaneously obtained when parameters are
modulated with smooth profiles. Our numerical results thus show that a
coupling scheme can at the same time be almost invisible when the
oscillator is in phase with its forcing cycle and effective enough to
ensure resetting when the oscillator is out of phase. By invisible, we
mean that the time profile remains in a close neighborhood of the
uncoupled one, so that the only effect of coupling is to fix the phase
of the oscillation with respect to the day/night cycle.

\subsection*{Robustness to daylight fluctuations}

Why would it be beneficial for a circadian oscillator to be minimally
affected by light/dark alternation in normal operation? A tempting
hypothesis is that while daylight is essential for synchronizing the
clock, its fluctuations can be detrimental to time keeping and that it
is important to shield the oscillator from them. If the entrained
temporal profile remains close to that of an uncoupled oscillator at
different values of the coupling parameter, then it will be naturally
insensitive to fluctuations in this parameter. To gain insight into
this fundamental question, we subjected the fully coupled and
occasionally coupled clock models to fluctuating daylight.

With the light input pathway unknown, we must allow for the fact that
light fluctuations may be strongly attenuated upon reaching the
\emph{Toc1}-\emph{Cca1} loop. For example, the light signal could be
transmitted through an ultrasensitive signaling cascade with almost
constant output above an input threshold close to daylight intensities
at dawn. The core oscillator would then be subjected to a driving
cycle much closer to a perfect square wave than the intensity profile.
We thus considered varying modulation depths for the core oscillator
parameters to reflect this possible attenuation.

\begin{figure}[!ht]
  \centering
  \includegraphics[width=15cm]{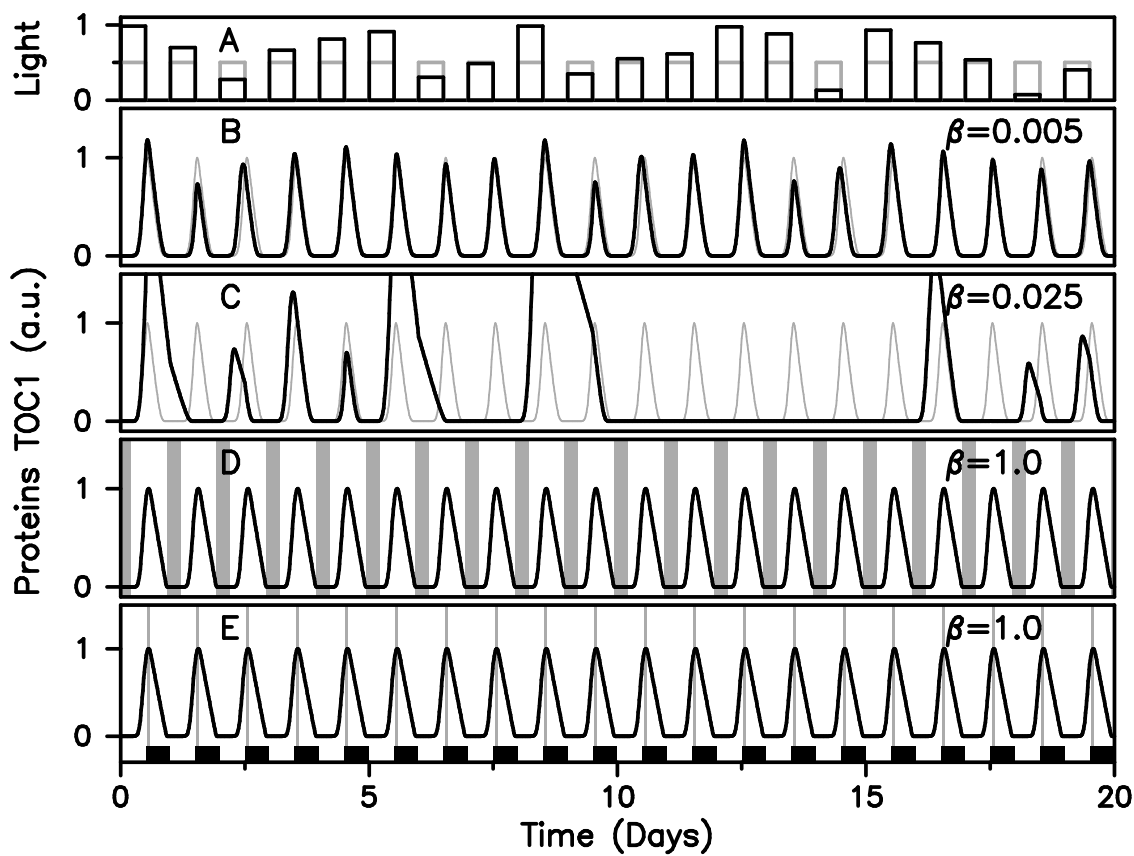}
  \caption{\textbf{Response of clock models to fluctuating daylight
      intensity}. (A) Light intensity varying randomly from day to
    day. The time evolution of TOC1 concentration is shown for: (B),
    (C) the permanently coupled clock model of
    Fig.~\ref{fig:adjustment}(A) at two different fluctuation levels,
    which are quantified by parameter $\beta$ (see Methods); (D) the
    clock model used in Fig.~\ref{fig:occas_coupling}(A)-(C); (E) the
    clock model used in Fig.~\ref{fig:occas_coupling}(D)-(F). When the
    clock operates nominally, numerical solutions (in black) and
    experimental time profiles (in gray) superimpose. }
  \label{fig:robustness}
\end{figure}

Although the two types of model adjust experimental data equally well
when subjected to a regular alternation, they have completely
different responses to daylight fluctuations. In
Fig.~\ref{fig:robustness}, we assume that light intensity is constant
throughout a given day but varies randomly from day to day. For almost
zero modulation, the fully coupled model of
Fig.~\ref{fig:adjustment}(B) maintains relatively regular
oscillations of varying amplitude (Fig.~\ref{fig:robustness}(B)). When
parameter values are modulated by only a few percent, however, this
model behaves erratically: oscillations stop for a few days,
expression peaks occur a few hours in advance,...
(Fig.~\ref{fig:robustness}(C)). A circadian clock similarly built
would be adversely affected by fluctuations in daylight intensity even
with very strong attenuation in the input pathway.

In contrast to this, the two occasionally coupled oscillators of
Fig.~\ref{fig:occas_coupling} keep time perfectly even for extreme
fluctuations (Figs.~\ref{fig:robustness}(D)-(E)) and generate
oscillations that are indistinguishable from those of the free-running
oscillator which adjusts experimental data recorded under strictly
periodic light/dark alternation.
Obviously, this extends to models combinining the two windows, such as
the one used in Fig.~\ref{fig:shiftrecover}. This simple model thus
describes a robust clock that is both sensitive to phase shifts in the
forcing cycle and insensitive to fluctuations in intensity.

We also studied the effect of fluctuations at shorter time scales.
When light intensity was varied randomly each hour, but with the same
mean intensity each day, the permanently coupled model was still
affected but much less than in Fig.~\ref{fig:robustness} (Fig. S6).

\subsection*{Influence of free-running period} 
\begin{figure}[htbp]
  \centering
    \includegraphics[width=14cm]{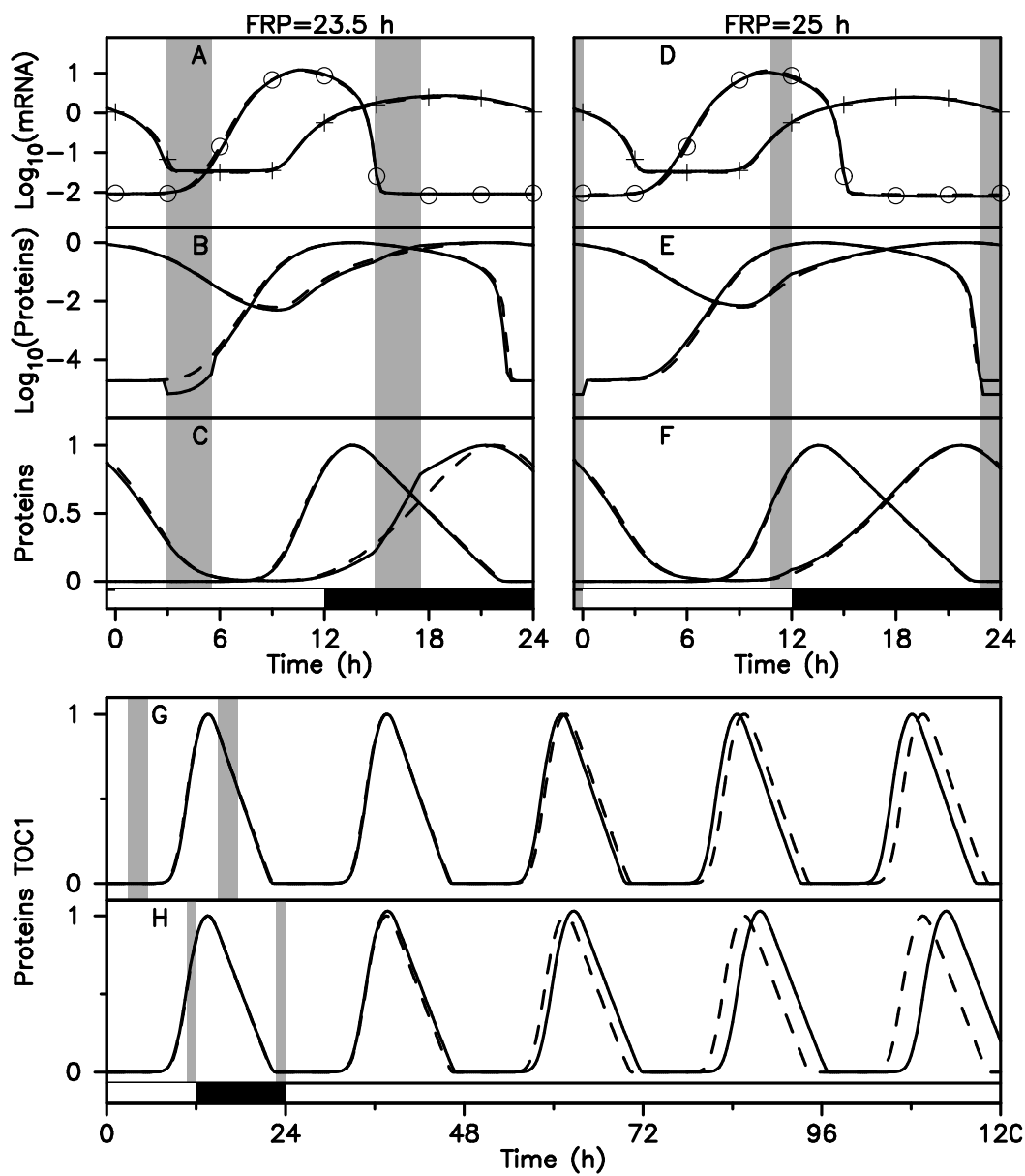}
    \caption{\textbf{Adjustment by models with gated coupling when FRP
        is different from 24 hours}. Gated coupling can also
      synchronize free-running clock models with a FRP of 23.5h or 25h
      without leaving any signature in mRNA profiles. Top left,
      (A)-(C): numerical solutions of model~(\ref{eq:model}) for a FRP
      of 23.5h, subjected to coupling windows shown as shaded areas.
      TOC1 (resp. CCA1) protein degradation rate is multiplied (resp.
      divided) by three from ZT3 to ZT5.5 (resp. ZT15 to ZT17.5). Top
      right, (D)-(F): numerical solutions of model~(\ref{eq:model})
      for a FRP of 25h, subjected to coupling windows shown as shaded
      areas. TOC1 (resp. CCA1) protein degradation rate is multiplied
      (resp. divided) by three from ZT22.75 to ZT24 (resp. ZT11.75 to
      ZT12). (A), (D) RNA in log scale; crosses (resp. circles)
      indicate \emph{Cca1} (resp. \emph{Toc1}) microarray data ; (B),
      (E) proteins in log scale; (C), (F) proteins in linear scale. In
      bottom panel, time evolution of TOC1 protein level (solid lines)
      during a transition from a 24-hour light/dark cycle to constant
      light compared to the forced profile (dashed line) for (G) a FRP
      of 23.5h and (H) a FRP of 25h.}
  \label{fig:frp}
\end{figure}

The results described
above may seem to rely on the FRP being equal to 24 hours. When the
FRP is smaller or larger, coupling is required to achieve frequency
locking and pull the oscillation period to 24 hours. To investigate
this more general case, we scaled kinetic constants of the
free-running model used in Fig.~\ref{fig:adjustment}(B) to shift the
FRP to 25 or 23.5 hours. In both cases (short FRP and long FRP), we
could find models with gated coupling that adjust perfectly the
experimental data with a period of 24 hours (Fig.~\ref{fig:frp}).
These models are very similar to those shown in
Fig.~\ref{fig:occas_coupling}, the only notable difference being that
coupling windows are shifted so that the induced resetting corrects
for the period mismatch. Interestingly, the coupling windows for a
FRP of 25 hours are located near the light/dark and dark/light
transitions.
We found that these coupling schemes were also very robust to daylight
fluctuations (Fig. S7), indicating that the modulation ratio (equal to
3 for the two windows) is
not critical.
We also found that without taking adjustment into account, the free
running oscillator is entrained by the coupling windows shown in
Fig.~\ref{fig:frp}) within a wide range of 
modulation ratios, from a lower threshold of 1.05 (resp. 1.25) for the
FRP equal 
to 23.5 hours (resp. 25 hours) to an upper threshold of 13 for both
FRPs. With a modulation ratio of 3, free-running oscillators with
FRPs ranging from 22 to 29 hours could be entrained. 

\subsection*{Gating by smooth profiles}
Gating of light input by rectangular profiles does not reflect the
fact that the concentration of the mediators modulating the oscillator
typically vary in a gradual way. The existence of nested coupling
windows such that models with shorter windows can adjust data with
larger parameter modulation (see Fig.~\ref{fig:rmserror}) suggests
investigating the action of smooth gating profiles, with maximal
parameter modulation near the center of the window. To this end, we
considered 24-hour periodic, Gaussian-shaped, modulation profiles
defined by: $1/r_C(t)=\delta^0_{P_C}/\delta_{P_C}(t)=
1+k_C\exp\left(-\frac{\sin(\pi(t-t_C)/24)^2}{\sigma_C^2}\right)$ and
$r_T(t)=\delta_{P_T}(t)/\delta^0_{P_T}=1+k_T\exp\left(-\frac{\sin(\pi(t-t_T)/24)^2}{\sigma_T^2}\right)$,
which are parameterized by the times of maximal modulation $t_{C}$,
$t_{T}$, the coupling durations $\sigma_C$, $\sigma_T$ and the
modulation depths $k_C$ and $k_T$. To assess whether good data
adjustment and resetting behavior could be obtained simultaneously,
these six parameters were chosen so as to minimize the RMS residual
phase error 5 days after an initial random phase shift ranging from
-12 to 12 hours (see Methods). Note that this naturally forces
adjustment to experimental RNA profiles.

\begin{figure}[htbp]
  \centering
  \includegraphics[width=15cm]{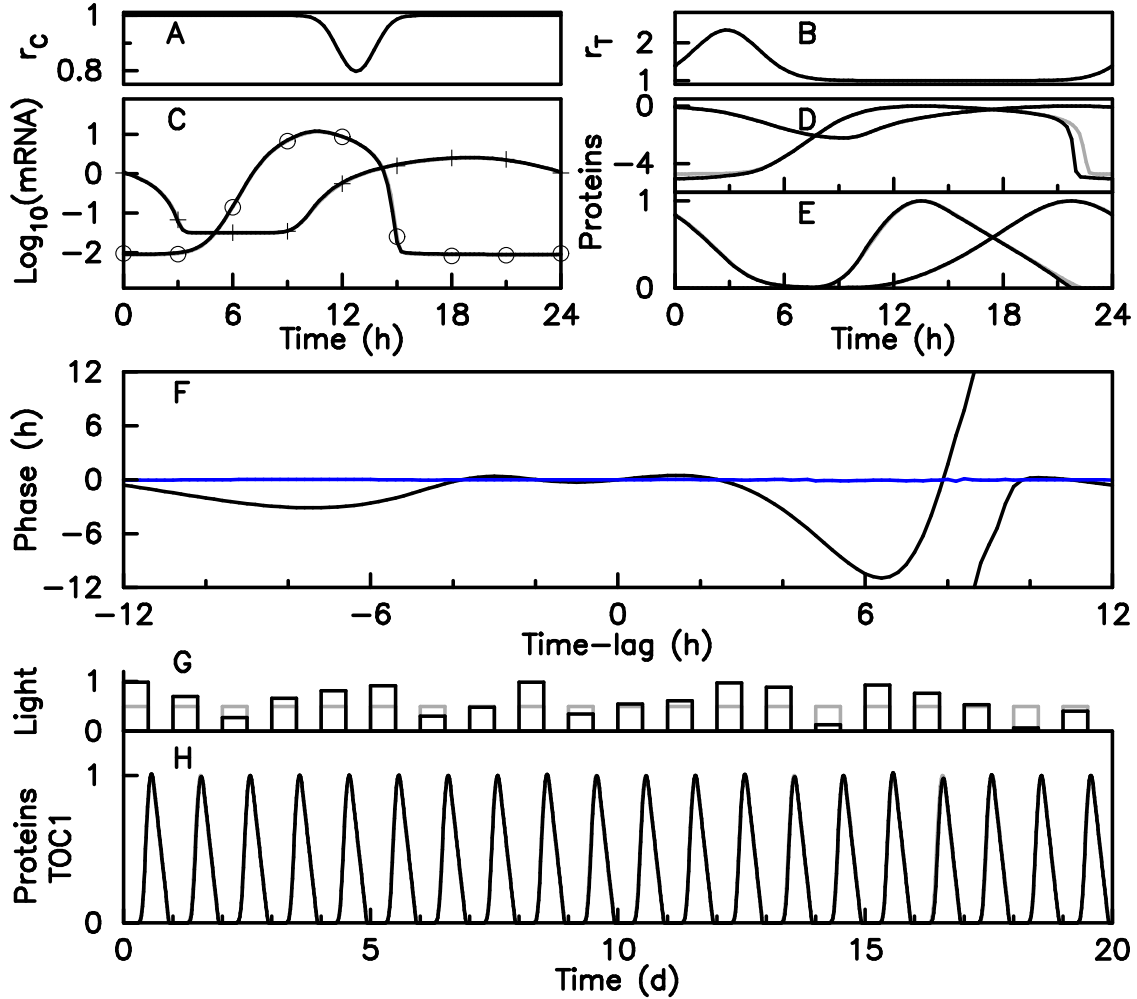}
  \caption{\textbf{Dynamical behavior of a clock model with gating by
      Gaussian-shaped modulation profiles}. (A) Temporal profile of
    the CCA1 protein stability modulation coefficient
    $r_C=\delta_{P_C}/\delta^0_{P_C}$ ($k_C=0.25$, $t_C=12.8$,
    $\sigma_C=0.17$) (B) Temporal profile of the TOC1 protein
    stability modulation coefficient $r_T=\delta_{P_T}/\delta^0_{P_T}$
    ($k_T=1.34$, $t_T=2.8$ and $\sigma_T=0.33$). (C), (D) and (E)
    display numerical solutions of model~(\ref{eq:model}) without
    coupling (in gray, same parameter values as in Fig~2(B)) and with
    coupling shown in (A) and (B) (in black). In (C) crosses (resp.
    circles) indicate the \emph{Cca1} (resp. \emph{Toc1}) microarray
    data used as target. Protein profiles are shown in (D)
    (logarithmic scale) and (E) (linear scale). (F): Resetting of the
    clock after a phase-shift of the day/night cycle. Solid curves
    display the residual phase shift of the clock after 1 (black) and
    5 (blue) day/night cycles as a function of the initial
    phase-shift. (G) Fluctuating daylight intensity (H) Response of
    the clock model with smooth coupling profiles to these
    fluctuations. The protein stability coefficients $k_X$ (see
    Methods) depend on daylight intensity $L \in [0,1]$ according to
    $k_X(L)=k_X(L=0.5)\times 3^{2L-1}$. }
  \label{fig:gaussian}
\end{figure}

The behavior of the model using the optimized modulation profiles
(Figs.~\ref{fig:gaussian}(A)-(B)) confirms the findings obtained with
rectangular profiles (Fig.~\ref{fig:gaussian}). The entrained RNA and
protein time profiles shadow that of the reference free-running
oscillator, with little evidence of the coupling
(Figs.~\ref{fig:gaussian}(C)-(E)). Phase resetting in response to a
phase shift is excellent (Fig.~\ref{fig:gaussian}(F)): RMS (resp.
maximum) residual phase shift after 5 days is 2.4 min (resp., 10 min).
This is all the more remarkable as the Gaussian shape of the
modulation profile is artificial, which shows that the dynamical
mechanism exploited here is robust and relatively insensitive to the
shape of the modulation profile. Moreover, the oscillator is extremely
resistant to daylight fluctuations (Fig.~\ref{fig:gaussian}(F)). In
spite of its simplicity, the two gene-oscillator studied here thus
fulfills key requirements for a circadian oscillator when modulated
with the right timing.

\section*{Discussion}

Our findings illustrate how mathematical modeling can give insight
into the architecture of a genetic module. Not only can expression
profiles of two \emph{Ostreococcus} clock genes be reproduced
accurately by a simple two-gene transcriptional feedback loop model,
but furthermore excellent adjustment of mRNA data is provided by a
free-running model. This counterintuitive result can be explained if
coupling to the diurnal cycle occurs during specific temporal windows,
where unidentified mediators interact with the TOC1-CCA1 oscillator in
such a way that it experiences negligible forcing when it is in phase
with the day/night cycle, and strong resetting when it is out of
phase. We could exhibit many coupling schemes compatible with
experimental mRNA temporal profiles, differing by the coupling
mechanism or by the window timing. This indicates that identification
of the actual light input pathway will require additional experimental
data. Our analysis strongly supports the conjecture that
\emph{Ostreococcus} genes \emph{Cca1} and \emph{Toc1} are the
molecular components of an oscillator at the core of
\emph{Ostreococcus} clock but does not exclude that other coupled
oscillators or feedback loops exist.

Why would a circadian oscillator decouple from the day/night cycle
when in phase with it so as to generate quasi-autonomous oscillations?
A natural hypothesis is that this protects the clock against daylight
fluctuations, which can be important in natural
conditions~\cite{Beersma08011999}. In a vast 
majority of numerical simulations and experiments on circadian clocks
reported in the literature, the day/night cycle is taken into account
through a perfect alternation of constant light intensity and
darkness. However, this is somehow idealized, as the primary channel
through which clocks get information about Earth rotation, namely
daylight, is variable.

In nature, the daylight intensity sensed by an organism depends not
only on time of day but also on various factors such as sky cover or,
for marine organisms such as \emph{Ostreococcus}, the distance to sea surface
and water turbidity, which can affect perceived intensity much more
than atmosphere. Therefore, the light intensity reaching a circadian
clock can vary several-fold not only from one day to the next but also
between different times of the day. A clock permanently coupled to
light is also permanently subjected to its fluctuations. Depending on
the coupling scheme, keeping time may become a challenge when
fluctuations induce phase resettings and continuously drive the
clock  away from its desired state.
Indeed, we found that a mathematical model with properly timed coupling windows
was insensitive to strong light intensity fluctuations while a
permanently coupled model became erratic even for very small coupling
strengths. 
For simplicity, we only tested the robustness of a model with
modulated TOC1 and CCA1 
protein degradation. However, it should be stressed that all other
light-coupling 
mechanisms that have been found to be robust with respect to adjustment
(see Figs.~\ref{fig:syst} and S4) are naturally also robust with
respect to daylight fluctuations. Indeed they adjust the experimental
data for varying coupling strengths at fixed window timings. This
indicates that the limit cycle is insensitive to variations in the
coupling strength, which is the key to the robustness to daylight
fluctuations.
Another interesting result from our numerical simulations is that the
most disruptive fluctuations are the variations in intensity from one
day to the other, since their time scale matches the oscillator
period. Indeed, faster or slower fluctuations are easily filtered out.

These results lead to enquire whether similar designs exist in
other circadian clocks.
Although the importance of this problem was noted some time
ago~\cite{Beersma08011999}, the robustness of circadian clocks to
daylight fluctuations and how this constraint shapes their molecular
architecture have been little studied until very
recently\cite{Troein20091961,Merrow2009R1042}. The discussion on how
genetic oscillators can keep daytime has essentially focused on the
most important sources of noise under constant conditions :
temperature
variations~\cite{Pittendrigh54:_comp_temp,Rensing02:_comp_temp,rand04:_desig}
or fluctuations in concentration due to small numbers of
molecules~\cite{DidierGonze01222002,barkai00:_circadiannoise}.
However, an operating clock is naturally subjected to an external
forcing cycle, which is yet another source of fluctuations.

We thus conjecture that a circadian clock must be built so as to be
insensitive to daylight intensity fluctuations when entrained by the
day/night cycle, just as it is insentitive to molecular or temperature
fluctuations, and that this can be achieved by keeping the oscillator
as close to the free-running limit cycle as possible, scheduling
coupling at a time when the oscillator is not responsive. An important
consequence of this principle is that it allows us to discriminate
between different possible coupling mechanisms for a given model, as
our analysis revealed dramatic differences in the ability of different
parametric modulations to buffer fluctuations. It also allows us to
determine the preferred timing for a given coupling mechanism, which
may prove very helpful when trying to identify the molecular actors
which mediate the light information to the clock.

When the FRP is close to 24 hours, as in much of our analysis it is
easy to understand why robustness to daylight fluctuations requires
that the forced oscillation shadows the free-running solution.
Robustness manifests itself in the time profile remaining constant
when subjected to random sequences of daylight intensity. This
includes strongly fluctuating sequences as well as sequences of
constant daylight intensity at different levels. Thus, the oscillator
response should be the same at high and low daylight intensities,
which implies that the solution must remain close to the free-running
one as forcing is increased from zero. Note that this only holds in
entrainment conditions, where coupling is not needed. When the clock
is out of phase, strong responses to forcing are expected, with
resetting being faster as forcing is stronger.

When the natural and external periods are significantly different, the
problem may seem more complex as coupling is required to correct the
period mismatch. There is a minimal coupling strength under which the
oscillator is not frequency-locked and entrainment cannot occur.
Nevertheless, we showed that properly timing the coupling windows is
as effective for oscillators with FRP of 23.5 and 25 hours as for the
24-hour example we had considered. Again, the forced solution remains
close to the free-running limit cycle even if proceeding at a
different speed to correct the period mismatch. This also shows that
FRP is not a critical parameter for adjustment of the experimental
data used here.

A consequence of the small deviation of the limit cycle from the
free-running one when coupling strength is varied is that oscillations
should vary little upon a transition from LD to LL or DD conditions
(see, e.g., Figs.~\ref{fig:frp}(G)-(H)). We 
searched the litterature for examples of such behavior.
Ref.~\cite{Leloup99a} provides a interesting comparison of models for
the \emph{Drosophila} and \emph{Neurospora} circadian clocks which is
illustrative for our discussion. In this study, the variation in
amplitude is much less pronounced for the \emph{Drosophila} model than
for the \emph{Neurospora} one (see Fig.~2 of~\cite{Leloup99a}).
Concurrently, the sensitivity of the phase of the entrained
oscillations to variations in the light-controlled parameter is much
smaller for the \emph{Drosophila} model (see Fig.~3 of
\cite{Leloup99a}), which is a necessary condition for robustness to
daylight fluctuations. Another interesting comparison involves the
one-loop and two-loop models of \emph{Arabidopsis}
clock\cite{locke05:_modelling,locke05:_extension}. The one-loop model
clearly modifies its behavior upon entering DD conditions from LD (see
Fig.~5 of~\cite{locke05:_modelling}) while the two-loop model
preserves its average waveform when transiting from LD to LL, except
for the disappearance of the acute response to light at dawn (see
Fig.~6 of \cite{locke05:_extension}). Thus,
the two-loop model not only reproduces experimental data better but
also seems more robust.

The \emph{Drosophila} and \emph{Neurospora} clock models analyzed
in~\cite{Leloup99a} also differ in their response to forcing when
their FRP is close to 24 hours~\cite{kurosawa06:_amplit}. A number of
circadian models cannot be entrained when their FRP is too close to 24
hours because complex oscillations, period-doubled or chaotic ones,
are observed easily for moderate to strong forcing. Indeed, it is
expected that near resonance between the forcing and natural periods,
the strong response exalts nonlinearities and favors complex behavior.
Again, the \emph{Drosophila} clock model appears to be more robust in
this respect~\cite{kurosawa06:_amplit}. We stress that making the
coupling invisible in entrainment conditions naturally addresses this
issue. Dynamically uncoupling the oscillator from the diurnal cycle in
entrainment conditions makes it immune both to fluctuations in
daylight intensity and to destabilization in the face of strong
forcing.

An important problem is how a clock with occasional coupling can
adjust to different photoperiods so as to anticipate daily events all
along the year. We can only touch briefly this question here as it
requires understanding how the temporal profile of the coupling
windows changes with 
photoperiod and thus a detailed description of the unknown light input
pathways and 
additional feedback loops that control the timing of these windows.
The key point is that the phase of the entrained oscillations is
controlled by the position of the coupling windows. Thus the role of
light input pathways and additional feedback loops, 
whose internal dynamics will typically be affected by input from
photoreceptors and feedback from the TOC1--CCA1 oscillator, is to time
the coupling windows as needed for each photoperiod so that the
correct oscillation timing is
generated~\cite{bognar99:_phytochrome,RekaToth12012001,salazar09:_predic}.
This 
question will be 
addressed in a future work, together with the analysis of the
luminescence time series recorded for differents photoperiods.

Our results also bring some insight into the recent observation that a
circadian clock may require multiple feedback loops to maintain proper
timing of expression peaks in response to noisy light input across the
year~\cite{Troein20091961}. We have shown here that a single two-gene
loop can display impressive robustness to daylight fluctuations when
its parameters are modulated with the right timing. As noted when
discussing the response to different photoperiods, this requires the
presence of additional feedback loops to generate the biochemical
signal needed to drive the core oscillator appropriately, and which we
have not yet identified and modeled in Ostreococcus. Robustness to
fluctuations thus implies a minimal level of complexity.

Finally, robustness to intensity fluctuations may explain why it is
important to have a self-sustained oscillator at the core of the
clock, as a forced damped oscillator permanently needs forcing to
maintain its 
amplitude, and is thereby vulnerable to amplitude fluctuations.
Confining the dynamics near the free-running limit cycle allows to
have a pure phase dynamics for the core oscillator, uncoupled from
intensity fluctuations. Understanding how to construct it will require
taking into account the sensitivity of the free-running oscillator to
perturbations across its cycle~\cite{D.ARand08062008}.

A simple organism as \emph{Ostreococcus} can apparently combine mathematical
simplicity with the complexity of any cell. The low genomic redundancy
of \emph{Ostreococcus} is certainly crucial for allowing accurate
mathematical modeling, leading to better insight into the clock
workings. \emph{Ostreococcus} therefore stands as a very promising model for
circadian biology, but also more generally for systems biology.

\section*{Materials and Methods}


  A minimal mathematical model of the transcriptional loop where
  \emph{Toc1} activates \emph{Cca1} which represses \emph{Toc1},
  consists of the following four differential equations:

  \begin{subequations}
    \label{eq:model}
  \begin{eqnarray}
    \dot{M_T} &=& \mu_T + \frac{\lambda_T}{1+(P_C/P_{C0})^{n_C}} -
    \delta_{M_T} \frac{K_{M_T} M_T}{K_{M_T} + M_T}\\
    \dot{P_T} &=& \beta_T M_T -
    \delta_{P_T} \frac{K_{P_T} P_T}{K_{P_T} + P_T}\\
    \dot{M_C} &=& \mu_C + \frac{\lambda_C (P_T/P_{T0})^{n_T}}{1+(P_T/P_{T0})^{n_T}} -
    \delta_{M_C} \frac{K_{M_C} M_C}{K_{M_C} + M_C}\\
    \dot{P_C} &=& \beta_C M_C -
    \delta_{P_C} \frac{K_{P_C} P_C}{K_{P_C} + P_C}
  \end{eqnarray}
\end{subequations}
Eqs~(\ref{eq:model}) describe the time evolution of mRNA
concentrations $M_C$ and $M_T$ and protein concentrations $P_C$ and
$P_T$ for the \emph{Cca1} and \emph{Toc1} genes, as it results from
mRNA synthesis regulated by the other protein, translation and
enzymatic degradation. \emph{Toc1} transcription rate varies between
$\mu_T$ at infinite CCA1 concentration and $\mu_T+\lambda_T$ at zero
CCA1 concentration according to the usual gene regulation function
with threshold $P_{C0}$ and cooperativity $n_C$. Similarly,
\emph{Cca1} transcription rate is $\mu_C$ (resp., $\mu_C+\lambda_C$)
at zero (resp., infinite) TOC1 concentration, with threshold $P_{T0}$
and cooperativity $n_T$. Translation of TOC1 and CCA1 occurs at rates
$\beta_T$ and $\beta_C$, respectively. For each species $Y$, the
Michaelis-Menten degradation term is written so that $\delta_Y$ is the
low-concentration degradation rate and $K_Y$ is the saturation
threshold.

Model~(\ref{eq:model}) has 16 free continuously varying parameters
besides the cooperativities $n_C$ and $n_T$ which can be set  to the
integer values 1 or
2 by the adjustment procedure.
mRNA 
concentrations are determined 
experimentally only relative to a reference value and protein profiles
are not adjusted. Therefore, two solutions of Eqs.~(\ref{eq:model})
that have the same waveforms up to scale factors are equivalent.
Therefore, we can eliminate four parameters by scaling
Eqs.~(\ref{eq:model}), with only 12 free parameters controlling
adjustment when parameters do not vary in time, which optimizes
parameter space exploration. Then parameters are rescaled so that the
maximum value of protein profiles is 100 nM, the maximum value of
\emph{Cca1} mRNA profile is 10 nM and the \emph{Toc1} and \emph{Cca1}
mRNA maximum values are in the same proportion as in microarray data.
This makes it easier to compare regulation thresholds and degradation
saturation thresholds relative to the maximum values of the four
concentrations. When the number of modulated parameters is $m$, parameter space is
$(12+m)$-dimensional.

Adjustment was carried out by using a large number of random parameter
sets as starting points for an optimization procedure based on a
Modified Levenberg--Marquardt algorithm (routine LMDIF of the MINPACK
software suite~\cite{more:_minpac}). Goodness of fit for a given
parameter set was estimated by the root mean square (RMS) error
between experimental and numerical mRNA levels, in logarithmic scale.
Numerical integration was performed with the SEULEX
algorithm~\cite{hairer96:_ODE}. Adjustment was carried out with 14
(resp. 2) Quad-Core Intel Xeon processors at 2.83 GHz during 72 hours
for the 28-dimensional (resp. 12-dimensional) parameter space.
Convergence was checked by verifying that the vicinity of the optimum
was well sampled. In the uncoupled case, the ODE system is invariant
under time translation so that its solutions are defined up to an
arbitrary phase. An additional routine was then used to select the
best-fitting phase.

To study the effect of daylight fluctuations, parameters were
modulated as follows. $L(t) \in \left[0,1\right]$ is the randomly
varying light intensity, with $L^{\text{ref}}=0.5$ the reference
level. We define the reference modulation depth of the $Y$ parameter
taking value $Y_L$ at standard light level and $Y_D$ in dark as
$k^{\text{ref}}_Y=\left(Y_L-Y_D \right)/\left(Y_L+Y_D \right)$. $L(t)$
modifies modulation depth according to
$k_Y=k_Y^{\text{ref}}\left[1 +
  \beta\;\left(L-L^{\text{ref}}\right)\right]$, where $\beta$
quantifies sensitivity to light variation. The modified modulation
depth fixes a new value for the day value, the dark value being
unchanged. For models with occasional coupling, we use similar
definitions with dark and light parameter values replaced by parameter
values respectively outside and inside of the coupling window. The
CCA1 stability modulation inside the window starting after dusk
depends on the intensity of the previous day.

The parameters of the Gaussian-shaped modulation profiles were
determined by optimizing resetting. For all possible variable initial
time lag ranging from -12 to 12 hours, the effect of the coupling
scheme based on the two profiles modulating TOC1 degradation and CCA1
degradation was characterized as follows. The time lag was applied to
the free-running cycle adjusting experimental data. Then, the coupling
scheme was applied for one or 5 days. Finally, the coupling was
switched off and the residual phase error was measured after two days.
The set of six parameters defining modulation profiles were obtained
as those which minimize RMS residual phase error across the 24-hour
interval. 

\section*{Acknowledgments} We thank Bernard Vandenbunder for his
helpful guidance in the early stages of this work oscillators and
Constant Vandermoere for 
assistance with data analysis.
\section*{References}

\clearpage
\section*{Figure Legends}

 \clearpage

\clearpage

\clearpage

\clearpage
\section*{Tables} 

\begin{table}[!ht]\centering \footnotesize
  
  \caption{\textbf{Model parameter values}}

    \begin{tabular}{|l|l|r|r|r|}\hline
      Symbol & Description  & FC
      (day) & FC (night) & FR      \\\hline

      $\mu_T$ & Minimal \emph{Toc1} transcription rate (nM/min) & 0.0017 & 0.0016 & 0.0065\\ 
      $\lambda_T$ & CCA1-dependent \emph{Toc1} transcription rate (nM/min) & 0.93& 0.29 & 0.67 \\
      $P_{C_0}$ & CCA1 level at \emph{Toc1} repression threshold (nM)& 1.47& 0.00& 1.04\\
      $n_C$ & Cooperativity of CCA1 &2&2&2\\
      $1/\delta_{M_T}$ & mTOC1 half-life (min)& 13.8& 22.0& 5.08  \\
      $K_{M_T}$ & mTOC1 degradation saturation threshold (nM)& 8.85& 18.3& 1.25\\
      $\beta_T$ & TOC1 translation rate (1/min)& 0.013& 0.023& 0.016\\
      $1/\delta_{P_T}$ & TOC1 half-life (min)& 29.9& 29.0 & 3.58 \\
      $K_{P_T}$ & TOC1 degradation saturation threshold (nM)& 3.85& 9.78& 0.76 \\
      $\mu_C$ & Minimal \emph{Cca1} transcription rate (nM/min) & 0.0075 &
      0.017 & 0.052 \\ 
      $\lambda_C$ & TOC1-dependent \emph{Cca1} transcription rate (nM/min) & 0.12 & 0.047& 0.060 \\
      $P_{T_0}$ & TOC1 level at \emph{Cca1} activation threshold (nM) & 100.4 & 1.49 & 44.1 \\
      $n_T$ & Cooperativity of CCA1 &2&2&2\\
      $1/\delta_{M_C}$ & mCCA1 half-life (min) &13.3 & 52.2&  0.82 \\
      $K_{M_C}$ & mCCA1 degradation saturation threshold (nM)& 0.56& 3.76& 0.063\\
      $\beta_C$ & CCA1 translation rate (1/min)& 0.056& 0.046& 0.075\\
      $1/\delta_{P_C}$ & CCA1 half-life (min)& 55.5& 92.3& 54.7 \\
      $K_{P_C}$ & CCA1 degradation saturation threshold (nM)&32.4 &
      36.0& 46.0\\ \hline
    \end{tabular}
    \label{tab:parameters}
  \begin{flushleft}
     Parameter values result
    from adjusting model 
    \eqref{eq:model} to experimental data (see Methods) with (i) all
    parameter values varying 
    between day and night (fully coupled model, FC, Fig. 2(A)) and (ii)
    all parameter values constant (free-running model, FR, Fig. 2(B)).
  \end{flushleft}
  \end{table}

  \clearpage

\renewcommand{\figurename}{Figure S}
\renewcommand{\tablename}{Table S}
\renewcommand{\caption}[1]{\parbox{\hsize}{\vspace{0.4cm}#1}\vspace{0.4cm}}

\centerline{\huge  \textbf{Supporting Information}}

\subsection*{Gated-coupling design in the weak modulation limit}%

The free oscillator model was shown to adjust remarkably well the
RNAmicroarray data from LD12:12 experiments.
A tempting hypothesis is that the synchronization of the free running
oscillator to the day-night cycle involves a light-dependent gated-coupling
mechanism that has restricted effect on the RNA traces when phase locked. 
We develop here a systematic method to repertoriate the coupling schemes 
that synchronize the free oscillator to the diurnal cycle while preserving the
adjustment score obtained in the absence of coupling. 
For enough weak coupling strength, any coupling schemes that achieve the
correct locking phase preserve the adjustement score.
Those coupling schemes can be found in the framework of perturbation theory
in the vicinity of a periodic orbit [1,2,3], assuming that the driving force
period is enough close to the internal clock period.
We consider the state vector of a nonlinear oscillator, which
represents the concentration of the molecular clock components. 
In constant dark conditions, the concentration vector $\mathbf{X}$  
evolves according to:  
\begin{equation}
d\mathbf{X}/dt=\mathbf{F}(\mathbf{X},\mathbf{p_0})
\label{eq:co}
\end{equation}
Eq. \ref{eq:co} has a periodic solution $\mathbf{X_{\gamma}}(t)$ corresponding
to a stable limit cycle of period $T$ close to $24$ hours.
We assume that the coupling between the light and the circadian oscillator is
mediated by a set of $N$ components ($k$ is the index), which modulate the
parameter vector in the direction of $\mathbf{dp}_k$:
\begin{equation}
\mathbf{p}(t)=\mathbf{p_0}+\sum_{k=1,N} L_k(t,\tau_k,(t_m)_k) \mathbf{dp}_k 
\end{equation}
where the $24h$-periodic scalar function $L_k(t,\tau_k,(t_m)_k)$ represents 
the temporal profile of activation (rectangular- or gaussian-shaped profiles
in the present paper) of the light-dependent component $k$ with $\tau_k$ and
$(t_m)_k$ characterizing the effective coupling window duration and center
($t=0$ correspond to the night-day transition or CT0).
  
A small enough parametric impulse perturbation applied at phase $u$
induces an infinitesimal change of the circadian oscillator phase defining a
$T$-periodic scalar function $Z_{piPRC}(u,\mathbf{dp})$  called infinitesimal
phase response curve [2] or, to be more precise, parametric impulse phase
response curve [3]:
\begin{equation}
Z_{piPRC}(u,\mathbf{dp}_k)=(\mathbf{dp}_k)^T.\mathbf{Z_p}(u)
\end{equation}
where
\begin{equation}
\mathbf{Z_p}(u)=[\frac{\partial \mathbf{F}(\mathbf{X}_{\gamma}(u))}
{\partial \mathbf{p}}]^T \frac
{\partial\phi(\mathbf{X}_{\gamma}(u))}{\partial{\bf X}}
\end{equation}

Then, the phase change induced by the light sensed during the daytime can
be derived from the convolution of the temporal profile of the light-sensing
components with the piPRC: 
\begin{equation}
\Delta\phi=\sum_{k=1,N} \int_0^{T} L_k(u,\tau_k,(t_m)_k)
Z_{piPRC}(u+\phi,\mathbf{dp}_k) du   
\end{equation}
where $\phi$ is the phase of the oscillator at CT0. 
A stable entrainment state requires that the scalar functions $L$ and
$Z_{ipPRC}$ satisfies:
\begin{equation}
\left\{ \begin{array}{rll}
\sum_{k=1,N} \int_0^{T} L_k(u,\tau_k,(t_m)_k) \,
Z_{piPRC}(u+\phi^*,\mathbf{dp}_k) du=\delta \phi^* \\  
\sum_{k=1,N} \int_0^{T} L_k(u,\tau_k,(t_m)_k) \,
Z_{piPRC}'(u+\phi^*,\mathbf{dp}_k) du<0 
\end{array} \right. 
\label{eq-gp}
\end{equation}
where $\phi^*$ is the locked phase (relative to CT0) and $\delta\phi^*$ is the
phase change 
induced by the period mismatch between the free oscillator and the day-night
period, which is assumed to be small with respect to $T$.

For any modulated parameter set $\mathbf{dp}$ whose $Z_{piPRC}$-function
is equal to $\delta \phi^*$, 
one can always find $\tau_k$ and $(t_m)_k$, that satisfies Eqs. \ref{eq-gp} 
above. 
In the case where there is a unique coupling scheme ($N=1$) with a rectangular
profile, the coupling interval satisfies:
\begin{equation}
\left\{ \begin{array}{rll}
\int_{t_m-\tau/2}^{t_m+\tau/2} \, Z_{piPRC}(u,\mathbf{dp}) du=\delta \phi^*\\
\int_{t_m-\tau/2}^{t_m+\tau/2} \, Z_{piPRC}'(u,\mathbf{dp}) du<0
\end{array} \right.
\label{eq-square}
\end{equation}
Figures 5 and S4 show the numerical solutions of this equation with $\delta
\phi^*$ equal to 0 (the FRP being equal to 24 hours), which determine the
coupling intervals (compatible with experimental data) for positive and
negative modulation of the 16 parameters of the model.

\section*{References}

\noindent
[1] Kramer MA, Rabitz H, Calo JM (1984) 
Sensitivity analysis of oscillatory systems. 
Appl Math Model. 8: 328-340.
\newline

\noindent
[2] Rand DA, Shulgin BV, Salazar D, Millar AJ (2004)
Design principles underlying circadian clocks.
J R Soc Interface. 1: 119-30. 
\newline

\noindent
[3] Taylor SR, Gunawan R, Petzold LR, Doyle FJ 3rd (2008)
Sensitivity Measures for Oscillating Systems: Application to Mammalian
Circadian Gene Network.
IEEE Trans Automat Contr. 53: 177-188.
\newline


\clearpage

\centerline{
  \includegraphics[width=12cm]{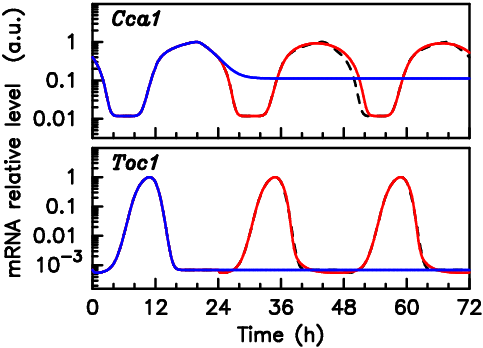}
}

\caption{Figure S1: \textbf{Transition from light/dark alternation
    (LD) to constant light (LL) and constant darkness (DD) for the
    fully coupled model.} Time evolution of mRNA concentrations for
  the fully coupled model shown in Fig.~2(A) for various light
  protocols: LD alternation (dashed, black), one LD period from ZT0 to
  ZT24 then constant light (in red) and one LD period from ZT0 to ZT24
  then darkness (in blue). \emph{Cca1} and \emph{Toc1} mRNA
  concentrations are shown in the top and bottom frame, respectively.}
    
\clearpage

\centerline{
  \includegraphics[width=10cm]{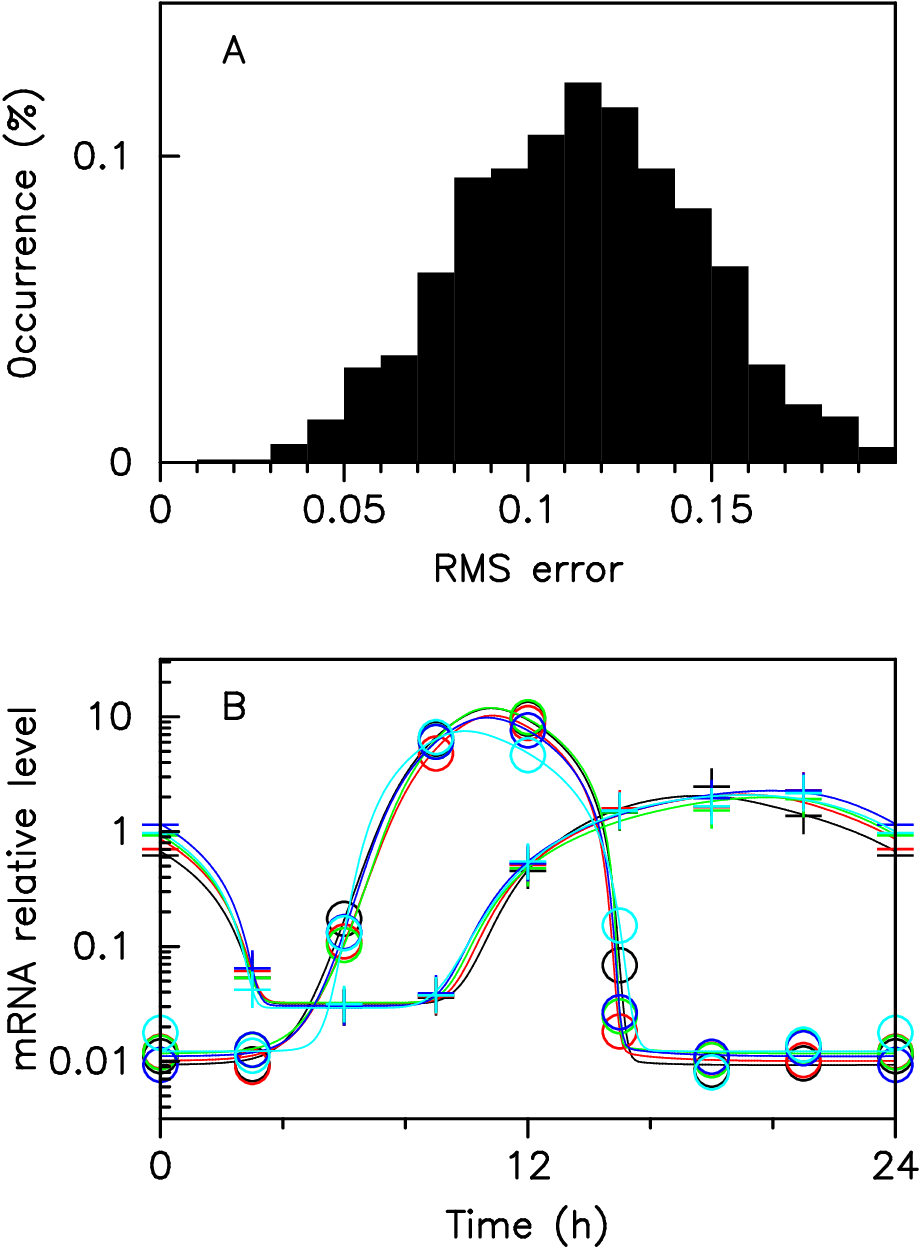}
}

\caption{Figure S2: \textbf{Influence of experimental errors on
    adjustement of a free running oscillator model to data}. Alternate
  target profiles with samples randomly chosen inside the interval of
  variation observed are generated and adjusted. Each random target
  corresponds to a slightly different parameter set and to a different
  adjustment RMS error (A) RMS error distribution; (B) The five target
  profiles most distant from each other have been selected and are
  associated with different colors. Crosses (resp. circles) indicate
  the \emph{Cca1} (resp \emph{Toc1}) mRNA target samples, the solid
  line is the numerical solution of the adjusting model.}

\clearpage

\centerline{
  \includegraphics[width=15cm]{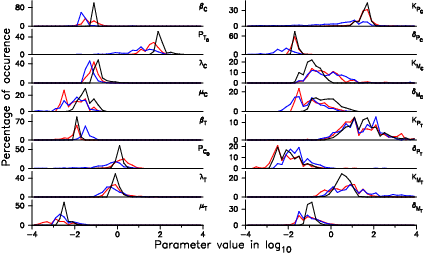}
}
    
    \caption{Figure S3: \textbf{Probability distribution for parameter values
      in parameter sets with adjustment RMS error below 10\%}.
      Parameters are determined as explained in Methods. The
      percentage of occurrence is evaluated for bins of width 0.2 in
      $\log_{10}$. The probability distributions of parameter values
      for the model with all parameters modulated are shown in red and
      blue for the day and night values, respectively. The probability
      distribution of parameter values for the model with all
      parameters constant is shown in black.}
\clearpage

\centerline{
  \includegraphics[width=11cm]{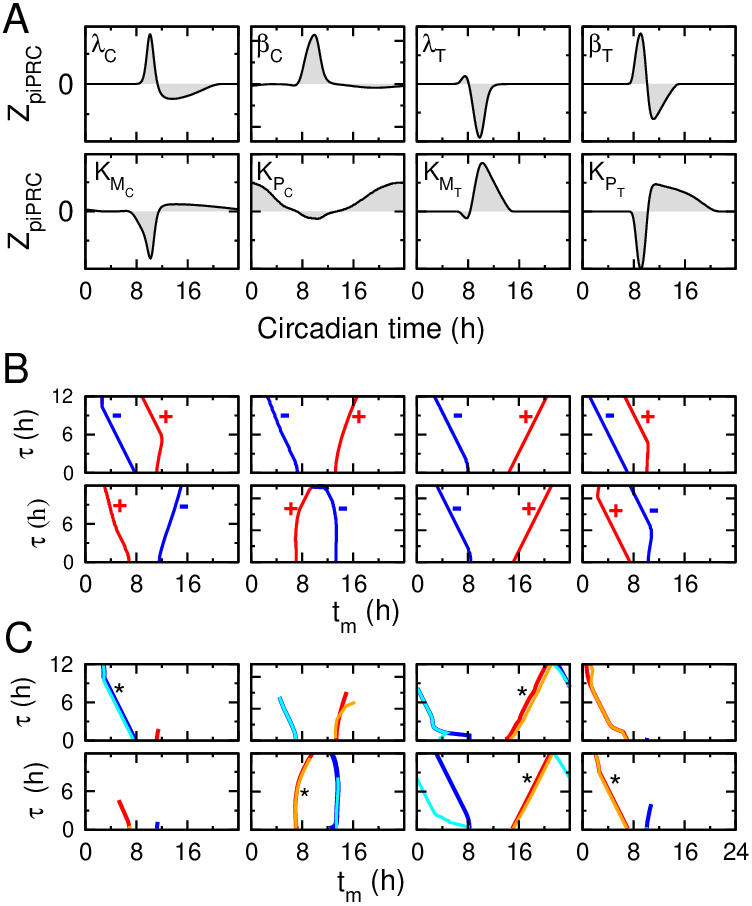}
}

\caption{Figure S4: \textbf{Characterization of
        coupling schemes.} (A) iPRC characterizing the phase change
      induced by an infinitesimal perturbation of parameters
      $\lambda_X$, $\beta_X$ and $K_X$. (B) Characterization of time
      position, $tm$, and duration $\tau$ of couplings with a
      rectangular gating profile satisfying Eq. (1). Parameters are
      modulated either positively (red) or negatively (blue). 
      (C) Characterization of time position and duration of couplings with
      a rectangular gating profile adjusting experimental data with a RMS
      error below $10\%$ for four different levels of coupling strength (blue:
      $p/p_0=1.17$; cyan: $p/p_0=1.17$; red: $p/p_0=0.85$; orange:
      $p/p_0=0.5$; $p/p_0$ being the ratio between the parameter values within
      and outside the coupling window)} 
\clearpage
\centerline{
  \includegraphics[width=11cm]{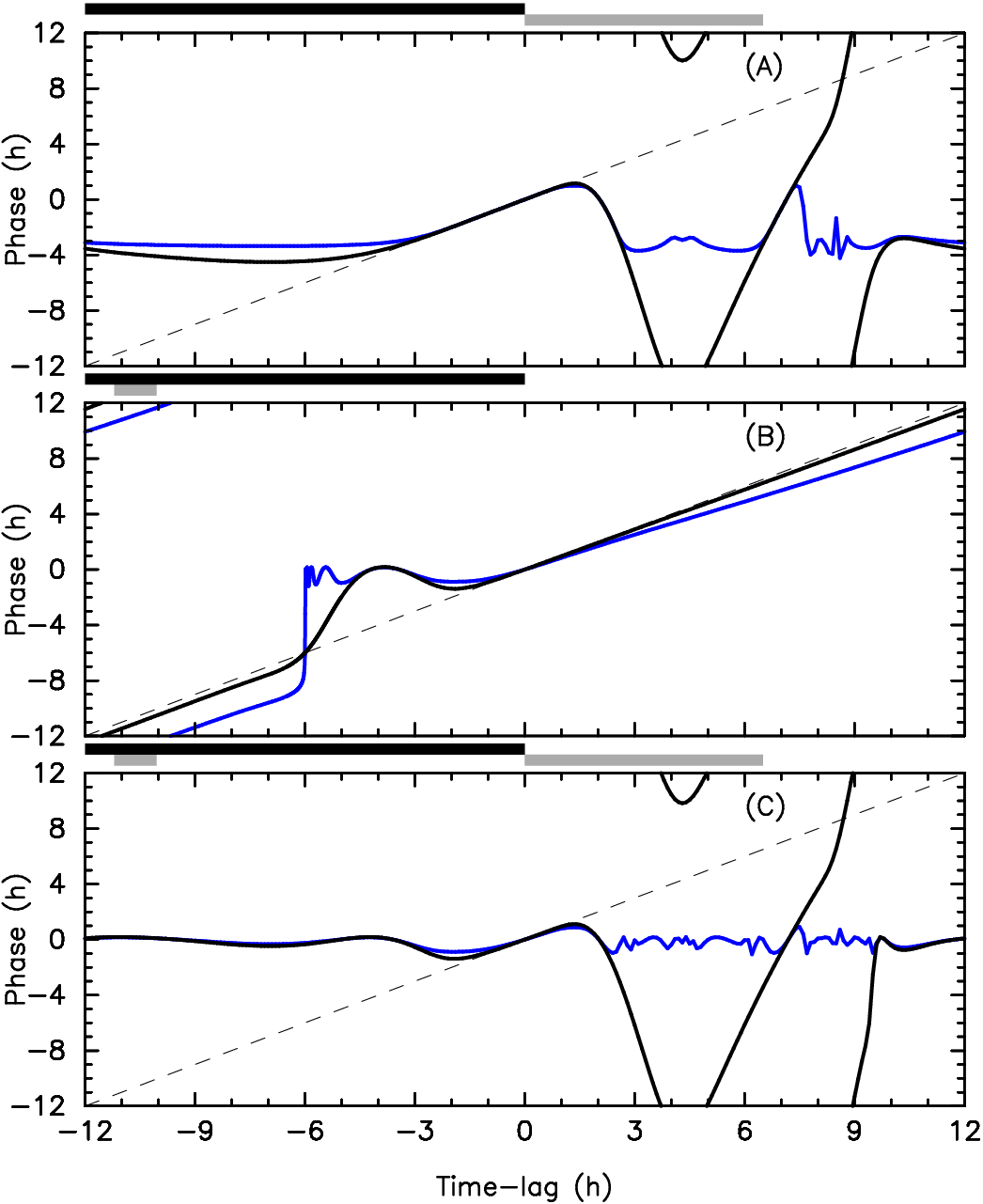}
}

\caption{Figure S5: \textbf{Resetting of the clock model of Fig.4 in
    response to a phase shift of the day/night cycle}. Solid curves
    display the residual phase shift of the clock after 1 (black) and
    5 (blue) day/night cycles as a function of the initial phase
    shift. (A) TOC1 degradation rate is multiplied by 2.1 between ZT0
    and ZT6.5. (B) CCA1 degradation rate is multiplied by 0.6 between
    ZT12.8 and ZT13.95. (C) Figure 6C is reproduced here for
    convenience. TOC1 (resp. CCA1) is multiplied by 2.1 (resp. 0.6)
    between ZT0 and ZT6.5 (resp. ZT12.8 and ZT13.95), which results in
    uniform convergence to phase-locking. Phase RMS error after 5
    day/night cycles is 25~min while the maximum error is 1~hour. }

\clearpage
\centerline{
  \includegraphics[width=15cm]{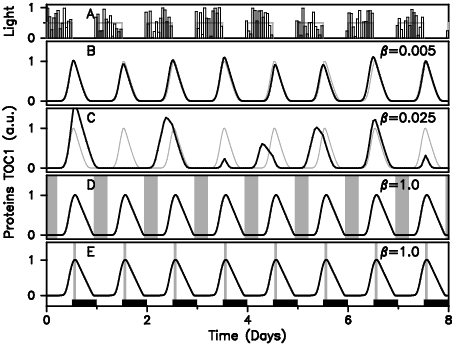}
}
\caption{Figure S6: \textbf{Response of the fully coupled and occasionally
    coupled clock models to fluctuations in daylight intensity
    occurring on a time scale of one hour}. The figure is otherwise
    similar to Fig~5. }

\clearpage
\centerline{
  \includegraphics[width=15cm]{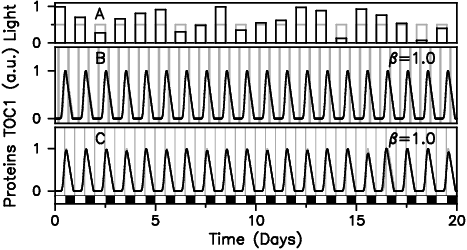}
}
\caption{Figure S7: \textbf{Response of the two occasionally coupled clock
    models of Fig.~8 to fluctuations in daylight intensity}. (a) Light
    intensity varying randomly from day to day. The time evolution of
    TOC1 protein concentration is shown for: (b) the clock model with
    a FRP of 23.5h; (c) the clock model with a FRP of 25h. The figure
    is otherwise similar to Fig~5.}

\end{document}